\documentclass[iop]{emulateapj}

\newcommand{\lya}{Ly-$\alpha$~}
\newcommand{\lyb}{Ly-$\beta$~}
\newcommand{\lyg}{Ly-$\gamma$~}

\newcommand{\nhi}{N_{\mhi}}

\newcommand{\nciv}{N_{\mciv}}
\newcommand{\pcmsq}{cm$^{-2}$}
\newcommand{\cii}{C~{\sc ii}~}
\newcommand{\ciii}{C~{\sc iii}~}
\newcommand{\civ}{C~{\sc iv}~}
\newcommand{\cv}{C~{\sc v}~}
\newcommand{\cvi}{C~{\sc vi}~}

\newcommand{\ovi}{O~{\sc vi}~}

\newcommand{\siiv}{Si~{\sc iv}~}
\newcommand{\hi}{H~{\sc i}~}

\newcommand{\heii}{He~{\sc ii}~}
\newcommand{\heiii}{He~{\sc iii}~}
\newcommand{\heiinsp}{He~{\sc ii}}
\newcommand{\heiiinsp}{He~{\sc iii}}
\newcommand{\civnsp}{C~{\sc iv}}

\newcommand{\ovinsp}{O~{\sc vi}}
\newcommand{\hinsp}{H~{\sc i}}
\newcommand{\ciiinsp}{C~{\sc iii}}
\newcommand{\ciinsp}{C~{\sc ii}}
\newcommand{\mgii}{Mg~{\sc ii}~}
\newcommand{\oden}{\rho/\bar{\rho}}

\newcommand{\mhi}{{\rm H \; \mbox{\tiny I}}}

\newcommand{\mciv}{{\rm C \; \mbox{\tiny IV}}}
\newcommand{\msiiv}{{\rm Si \; \mbox{\tiny IV}}}

\newcommand{\mheii}{{\rm He \; \mbox{\tiny II}}}

\newcommand{\kms}{km s$^{-1}$}

\begin{document}

\title{The Carbon Content of Intergalactic Gas at $z=4.25$ \\
  and its Evolution Toward $z=2.4$\altaffilmark{1}}

\author{Robert A. Simcoe\altaffilmark{2,3}}

\altaffiltext{1}{This paper includes data gathered with the 6.5 meter
  Magellan Telescopes located at Las Campanas Observatory, Chile}
\altaffiltext{2}{MIT-Kavli Center for Astrophysics and Space Research}
\altaffiltext{3}{Alfred P. Sloan Research Fellow}

\begin{abstract}

This paper presents ionization-corrected measurements of the carbon
abundance in intergalactic gas at $4.0 < z < 4.5$, using spectra of
three bright quasars obtained with the MIKE spectrograph on Magellan.
By measuring the \civ strength in a sample of 131 discrete
\hinsp-selected quasar absorbers with $\oden \ge 1.6$, we derive a
median carbon abundance of [C/H]$=-3.55$, with lognormal scatter of
approximately $\sigma\approx 0.8$ dex.  This median value is a factor
of two to three lower than similar measurements made at $z\sim 2.4$
using \civ and \ovinsp.  The strength of evolution is modestly
dependent on the choice of UV background spectrum used to make
ionization corrections, although our detection of an abundance
evolution is generally robust with respect to this model uncertainty.
We present a framework for analyzing the effects of spatial
fluctuations in the UV ionizing background at frequencies relevant for
\civ production.  We also explore the effects of reduced flux between
3-4 Rydbergs (as from \heii Lyman series absorption) on our abundance
estimates.  At \heii line absorption levels similar to published
estimates the effects are very small, although a larger optical depth
could reduce the strength of the abundance evolution.  Our results
imply that $\sim 50\%$ of the heavy elements seen in the IGM at $z\sim
2.4$ were deposited in the 1.3 Gyr between $z\sim 4.3$ and $z\sim
2.4$.  The total implied mass flux of carbon into the \lya forest
would constitute $\sim 30\%$ of the IMF-weighted carbon yield from
known star forming populations over this period.

\end{abstract}


\section{Introduction}\label{sec:intro}

Intergalactic \civ has been studied in the spectra of quasars at
essentially all redshifts, from the very local neighborhood in the
observed-frame UV \citep{cooksey_civ} to redshift six in the
observed-frame near-infrared \citep{becker_civ, ryan_weber_civ,
  simcoe_z6, ryan_weber_1}.  Yet most of our detailed knowledge about
the ionization-adjusted metal content of intergalactic matter comes
from studies at $z\sim 3$ \citep{schaye_civ_pixels, simcoe2004}.
In this interval the observed-frame transitions of both \civ and \lya
fall in an optimal region for optical spectrometers, in terms of both
sensitivity and lack of foreground contaminants.

However, even at $z\sim 3$ the density of intergalactic gas is
sufficiently low, and the UV ionizing background field is sufficiently
intense, that most carbon is ionized to the \cv state and higher. At
typical IGM densities for $z\sim 3$, about 2\% of all carbon atoms
are in the \civ state.  This fact, coupled with the low value of both
the particle density and the overall heavy element abundance, leads to
an extremely weak \civ signal in the more tenuous regions of the
cosmic web.  Consequently studies of metal abundances have required
extremely high signal-to-noise ratio data ($\sim 50-200$).

To date the most sensitive surveys have either utilized statistical
methods to tease signal from the distribution of \civ and \hi pixel
optical depths \citep{schaye_civ_pixels, ellison_civ,
  songaila_new_civ}, or used the \ovi line which has a higher
ionization potential (and therefore a stronger signal) to trace
metallicity \citep{simcoe2004,dave_ovi, bergeron_ovi}.  When the same
UV backgrounds are used to estimate ionization corrections, both the
\civ pixel method and the \ovi method yield an IGM abundance of
roughly [C/H]$=-3.1$ or [O/H]$=-2.8$ at $z\sim 3$.

The \ovi line does not lend well to studies of abundance evolution at
high redshifts, because it is blended in the \lya forest, whose
overall opacity increases at higher $z$.  Above $z\sim 3$ it
becomes effectively impossible to deblend \ovi doublets from \hi
forest lines, even in high resolution spectra.  \civ {\em is} seen at
these redshifts, with a distribution of strengths that is nearly
independent of redshift \citep{songaila_new_civ}.  

\begin{figure*}
\epsscale{1.0}
\plotone{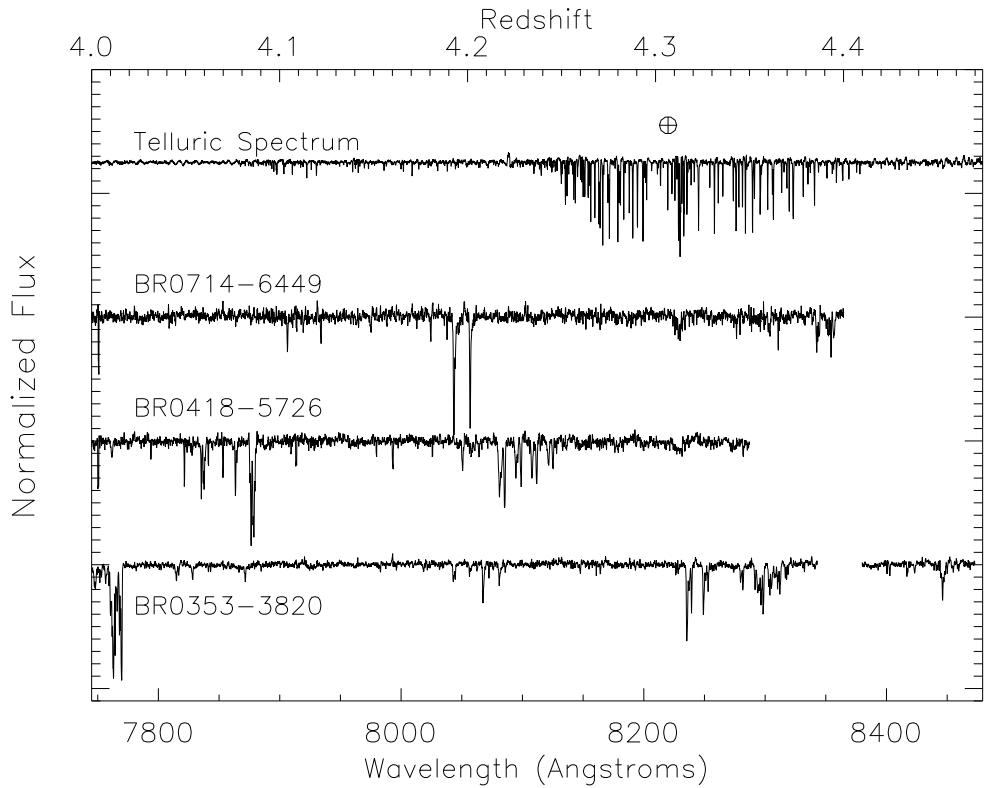}
\caption{MIKE spectra of the \civ region for the three objects used
  for the $z\sim 4.3$ sample.  At top, the mean telluric absorption
  spectrum determined by standard star observations is shown for
  reference.  The masked region at 8350\AA in the spectrum of BR0353
  is at the location of a known \mgii interloper.}
\label{fig:spectra}
\end{figure*}

\citet{schaye_civ_pixels} systematically examined the ionization corrected
carbon abundance between $3.0 < z < 4.0$, finding no statistically
significant evidence of evolution.  However the bulk of this sample
was centered at lower redshift, with only one object at $ > 4$.  

In this paper, we revisit the question of abundance evolution by
obtaining three new high-SNR spectra of $z\sim 4.5 $ quasars.  We use
these to make a detailed comparison of \civ absorption at $z\sim 4.3$
with the \ovi and \civ measurements we obtained earlier at $z\sim
2.5$.

The $z\sim 4.0-4.5$ range is a ``sweet-spot'' for observing \civ
because it is the interval when the fraction of carbon in the triply
ionized state peaks over cosmic time.  As one moves to higher
redshifts, the predominant ionization state moves to progressively
lower levels due to the combined $(1+z)^3$ increase in the baryon
density and decline of the hard-UV radiation field provided by
quasars.  Thus a survey of {\em individual} \civ systems at $z\sim
4.3$ is comparably sensitive to a survey for \ovi at $z\sim 2.5$ or a
statistical detection of \civ at $z\sim 3$.  These different surveys
all probe gas structures with baryonic overdensities of $\oden\sim 1.5$
relative to the cosmic mean at their respective epochs.

In Section \ref{sec:observations} we describe the new observations and
methods for data processing; Section \ref{sec:sample} describes our
sample selection, and Section \ref{sec:km_lines} describes our observational
measurements.  Section \ref{sec:uvbg} discusses ionization corrections
applied to derive the abundance measurements presented in Section
\ref{sec:abundance}.  The resultant cosmic abundance evolution and
implications for galaxy formation are discussed in Section
\ref{sec:discussion}.  Throughout we assume a cosmology with
$\Omega_M=0.3, \Omega_\Lambda=0.7$, and $H_0=71$ km/s/Mpc.

\section{Observations}\label{sec:observations}

We observed the three bright, southern quasars listed in Table 1 using
Magellan/MIKE over several different observing runs between 2004-2006.
Spectra were obtained using a 0.875\arcsec slit, which yields a
combined velocity resolution of 14\kms, as measured from concurrent
exposures of ThAr arc lamps.  Extraction was performed using a
customized software package for MIKE \citep{bernsteinMIKE}, which
makes use of the Poisson-limited 2D sky subtraction methods outlined
in \citet{kelson}.  The Echelle orders of individual exposures were
flux calibrated using observations of hot spectrophotometric standard
stars.  Finally, the calibrated orders from all exposures for each
object were combined onto a single 1-D wavelength grid and converted
to vacuum-heliocentric units.

\begin{deluxetable}{c c c c}
\tablewidth{0pc}
\tablecaption{$z\sim 4.5$ MIKE Targets}

\tablehead{{Object} & {$r$} & {$z_{min}$} & {$z_{max}$} }
\startdata
BR0353$-$3820 & 18.0 & 4.024  & 4.475\\
BR0418$-$5726 & 17.8 & 4.000  & 4.353\\
BR0714$-$6449 & 18.4 & 4.122  & 4.397\\
\enddata

\end{deluxetable}

\begin{figure*}
\epsscale{1.0}
\plottwo{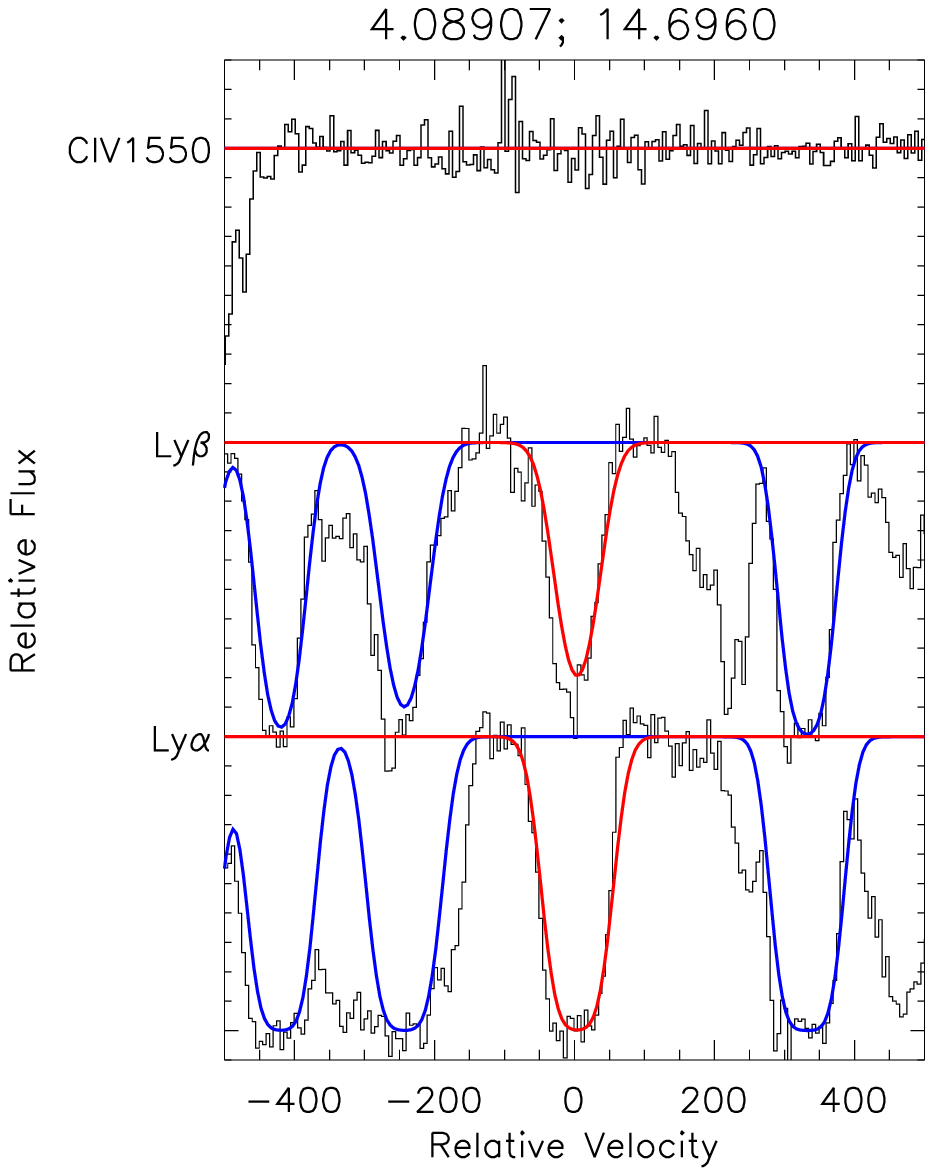}{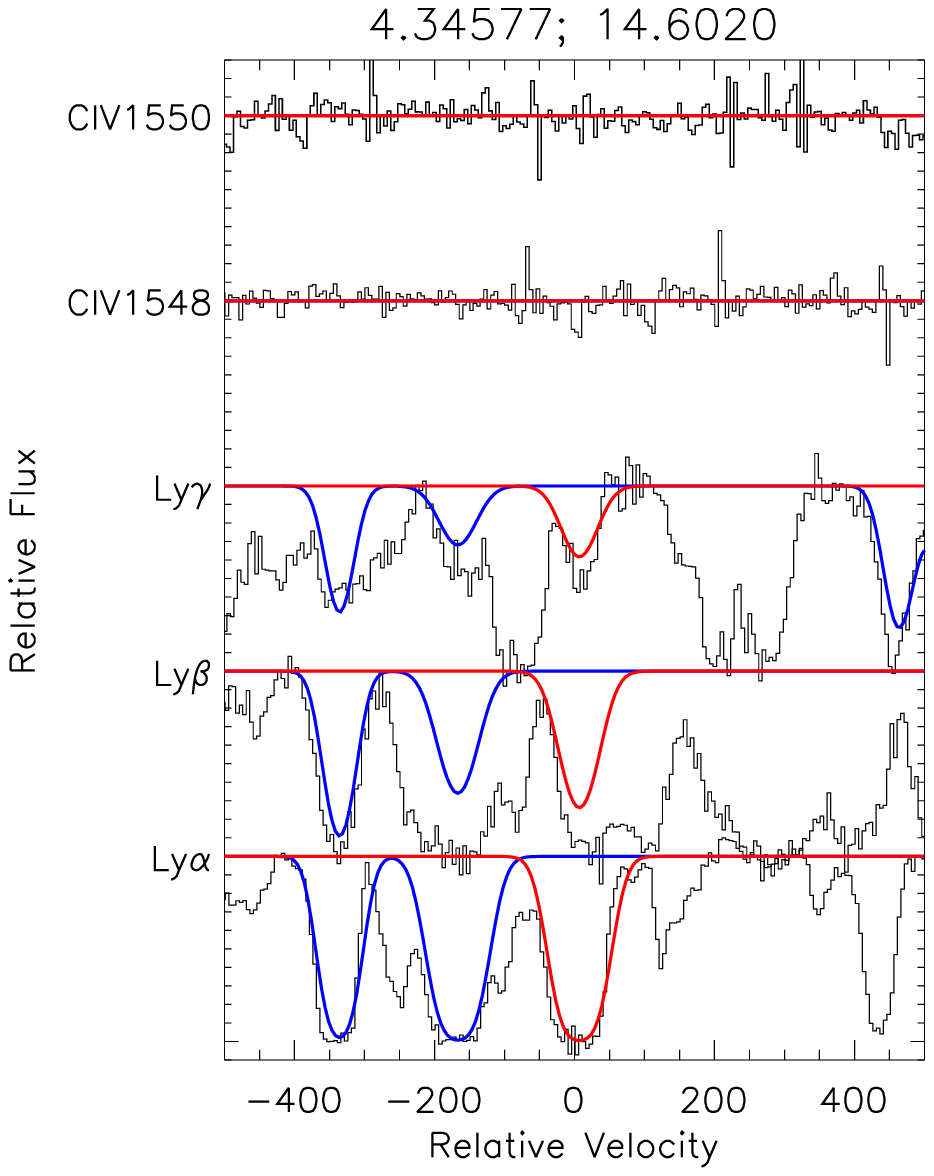}
\plottwo{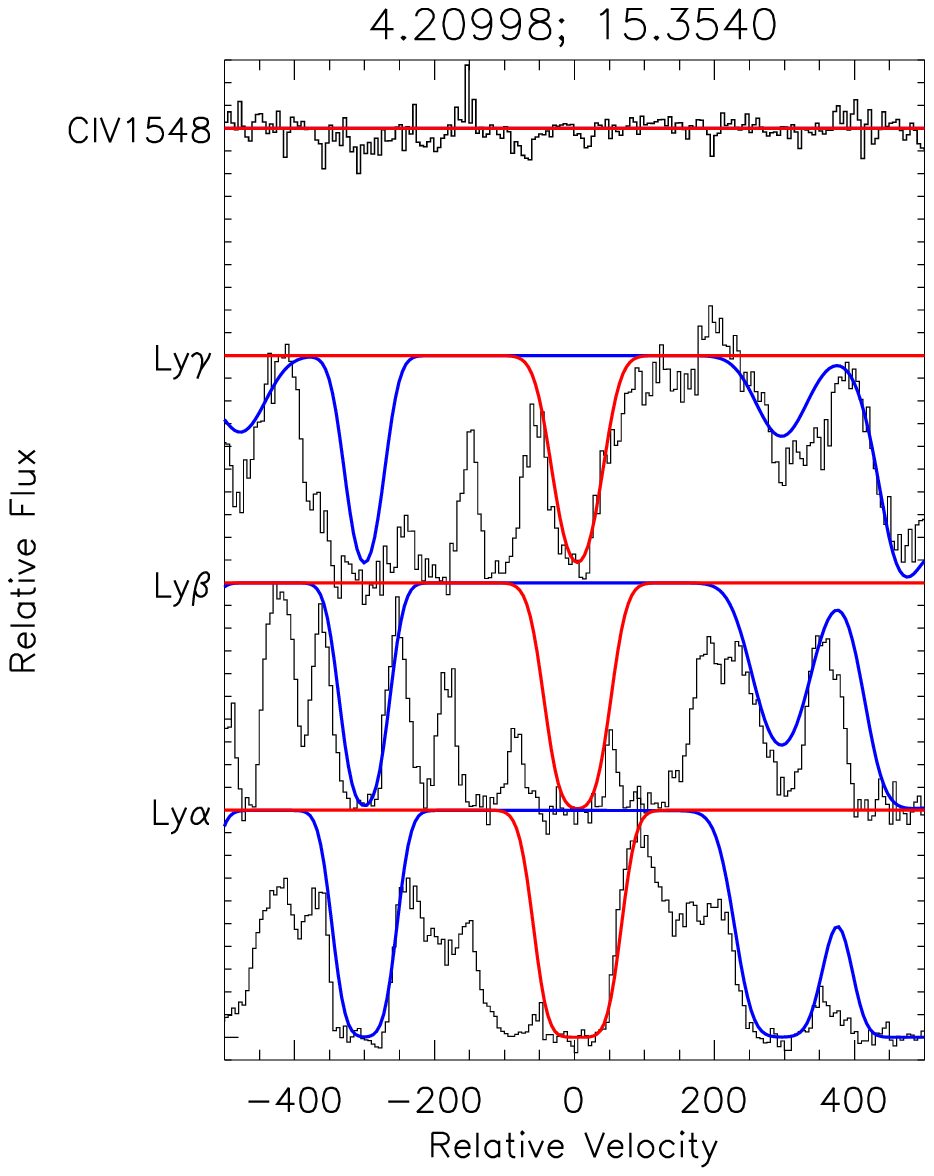}{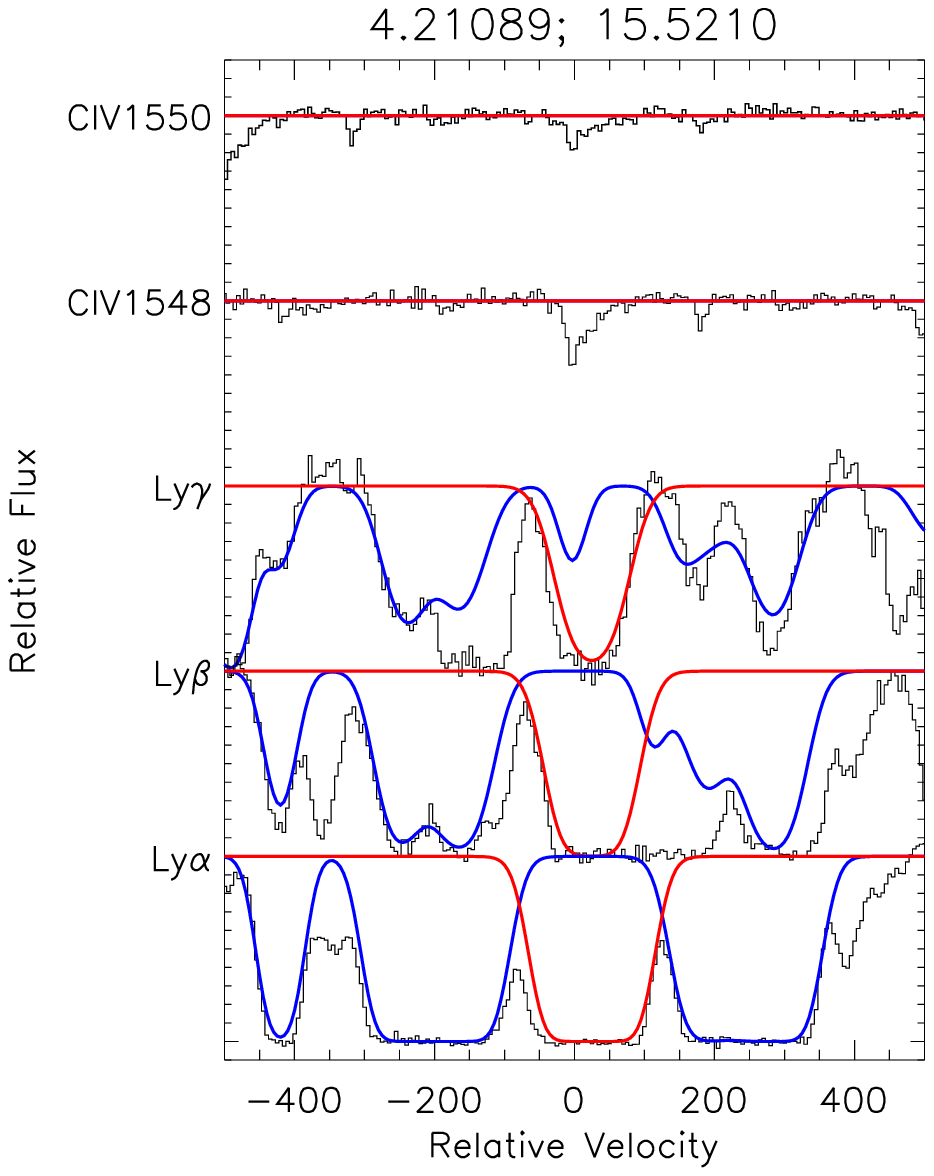}
\plottwo{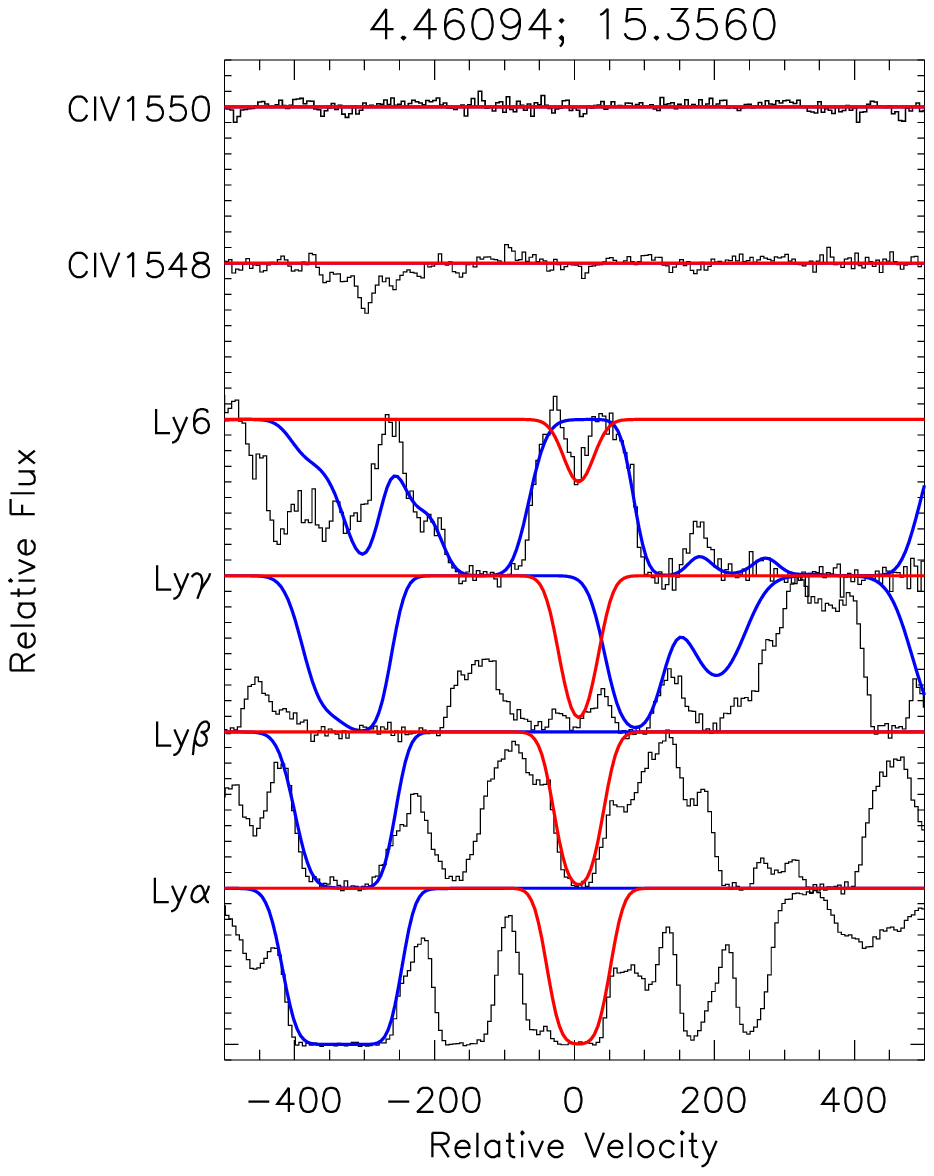}{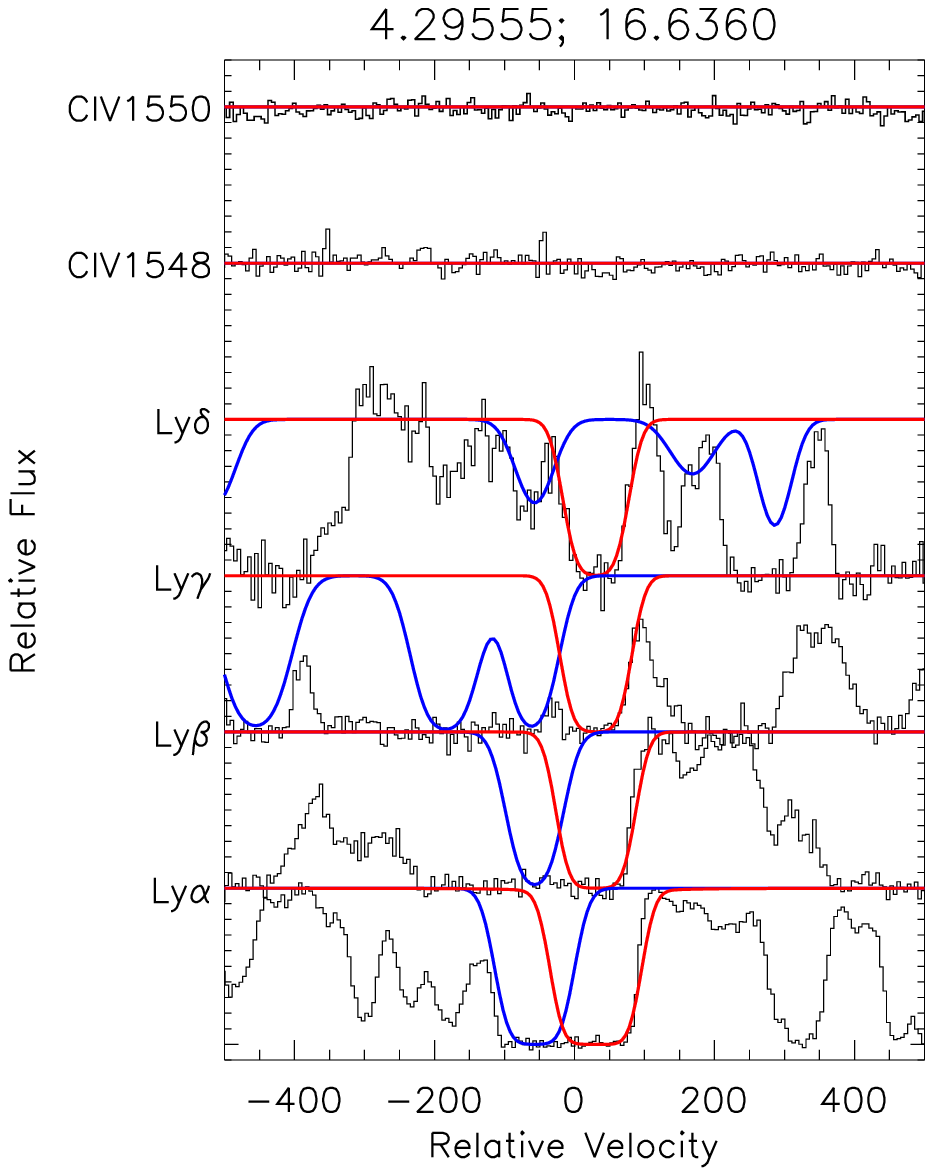}
\caption{Montage of absorbers and corresponding best-fit Voigt
  profiles, illustrating the use of partial profiles for multiple
  Lyman series transitions.  For each plot, the column density and
  redshift of the line shown in red are given at top.  Blue curves
  show blended absorption for other lines with $\nhi>10^{14.5}$, which
    also are included in the \civ sample.}
\label{fig:systems}
\end{figure*}

At $z=4.0-4.5$, the \civ transition falls between $\lambda=$7745\AA
~and 8520\AA.  This region is rich with telluric absorption features
from the atmospheric ``A'' and ``B'' bands of molecular oxygen and
water.  We carefully removed these features from our QSO spectra by
fitting transmission functions to observations of our standard stars
interspersed throughout each night.

The final 1-D spectra are shown in Figure \ref{fig:spectra}.  The
resultant signal-to-noise ratio in the \civ region ranges from 20 to
40; in the \lya forest the ratio is somewhat lower but still suitable for
measuring \hi given the much greater strength of the features being
measured.

\section{Sample Selection}\label{sec:sample}

To choose our spectral search regions for \civnsp, we began by fitting
Voigt profiles to the entire \hi \lya forest using the {\tt
  vpfit}\footnote{http://www.ast.cam.ac.uk/$\sim$rfc/vpfit.html} software
package.  We selected $\nhi \ge 10^{14.5}$ as the minimum threshold
column density for an absorber to be included in the \civ candidate
sample.  For a standard \citet{haardt_cuba} model of the UV background
field, an absorber with this column density and [C/H]=-3.0 would have
$\nciv=10^{12}$\pcmsq---roughly the detection limit of our data.

At $z\sim 4.3$, this \hi column density corresponds to a baryonic
overdensity of $\oden=1.6$ relative to the cosmic mean
\citep{schaye_forest}.  Coincidentally, this threshold is nearly
identical to the minimum $\oden$ where individual \ovi lines can be
detected at $z\sim 2.5$; \civ lines can only be seen at higher
overdensity at $z\sim 2.5$.  Throughout the present paper we quote
comparisons to both \ovi and \civ samples at lower redshift.  

\subsection{Reliability Tests for $N_\mhi$ Fits}

When fitting for $\nhi$ at $z > 4$, much of the Lyman series is
accessible provided there are no strong Lyman limit systems.  We used
at least three transitions of the Lyman series, and often 4 or 5 to
constrain the column densities of each \hi line.  If no unsaturated
Lyman series line or profile wing was available, a line was discarded
from the sample.

\begin{figure}
\epsscale{1.0}
\plotone{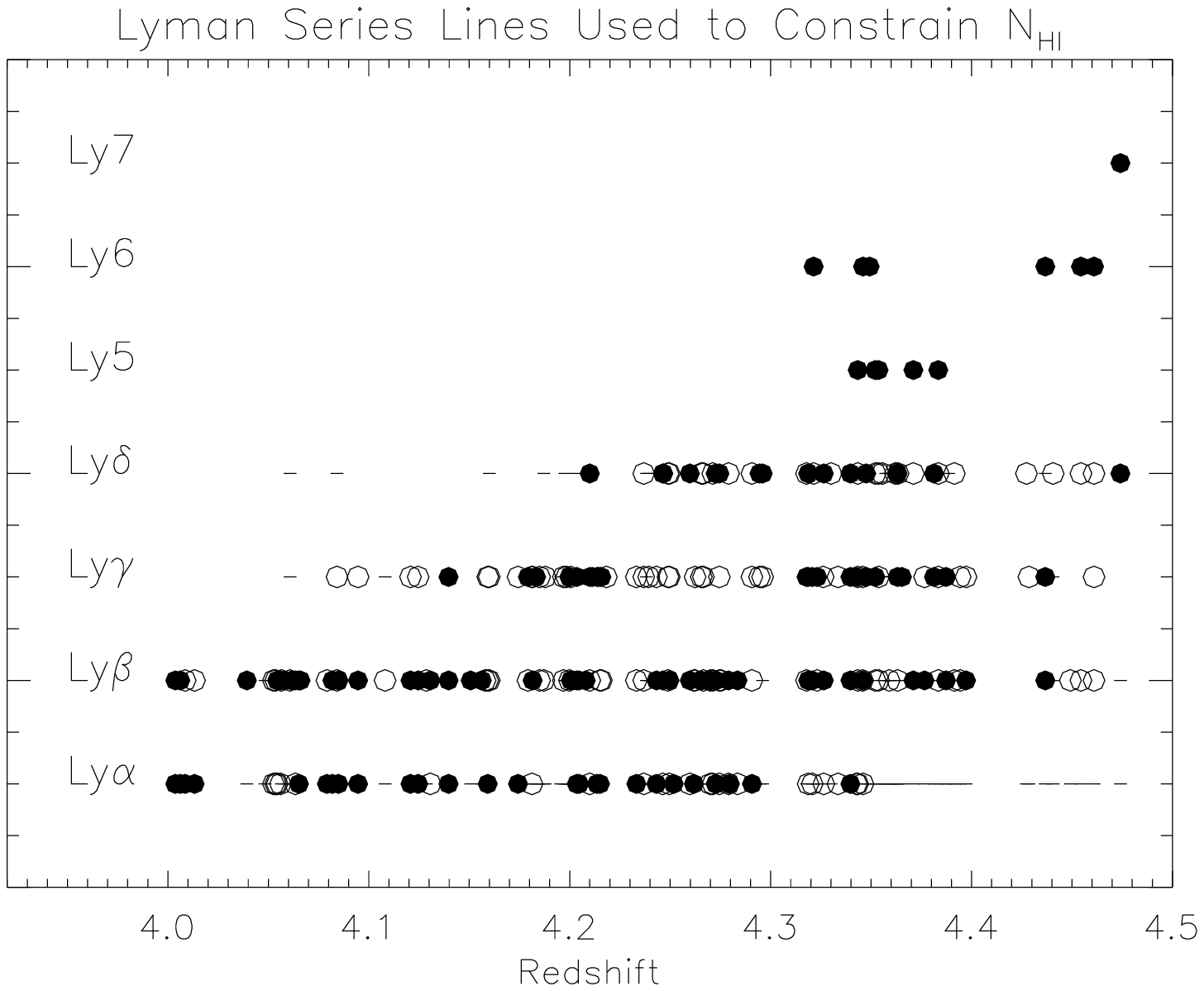}
\caption{Illustration of which Lyman series lines were used in \hi
  fits at varying redshifts.  For each system, multiple transitions
  are used.  Horizontal lines indicate components with at least
  partial saturation (which sometimes contain information in
  absorption wings).  Open circles are transitions which may be
  blended but are not saturated in absorption and hence provide upper
  limits on $\nhi$.  Solid circles indicate distinct, apparently
  unblended lines which align in velocity with other Lyman series
  absorption.}
\label{fig:lyseries}
\end{figure}

Figure \ref{fig:systems} shows several examples of \hi absorbers fit
in this manner, selected to span the range of column densities in our
sample.  Clearly there is significant blending of lines in the \lya
and higher order transitions, and it is rare to have a single absorber
with many unblended transitions.  However roughly 60\% of the sample
has at least one Lyman series transition completely free of blending,
and an additional 20\% were significantly constrained by the
unsaturated wings of blended lines in the high order transitions.
Figure \ref{fig:lyseries} shows a graphical summary of which
transitions were used to constrain each line in the sample, over the
full range in redshift.

In practice, very few systems were rejected because of saturation in
all Lyman series lines.  
For each sightline we truncated our \civ search below the redshift where
the \lyg line becomes saturated from Lyman limit (or completely
blended Lyman series) absorption.  The high redshift boundary for each
sightline is located $5000$ \kms blueward of the QSO's emission
redshift.

\begin{figure}[b]
\epsscale{1.0}
\plotone{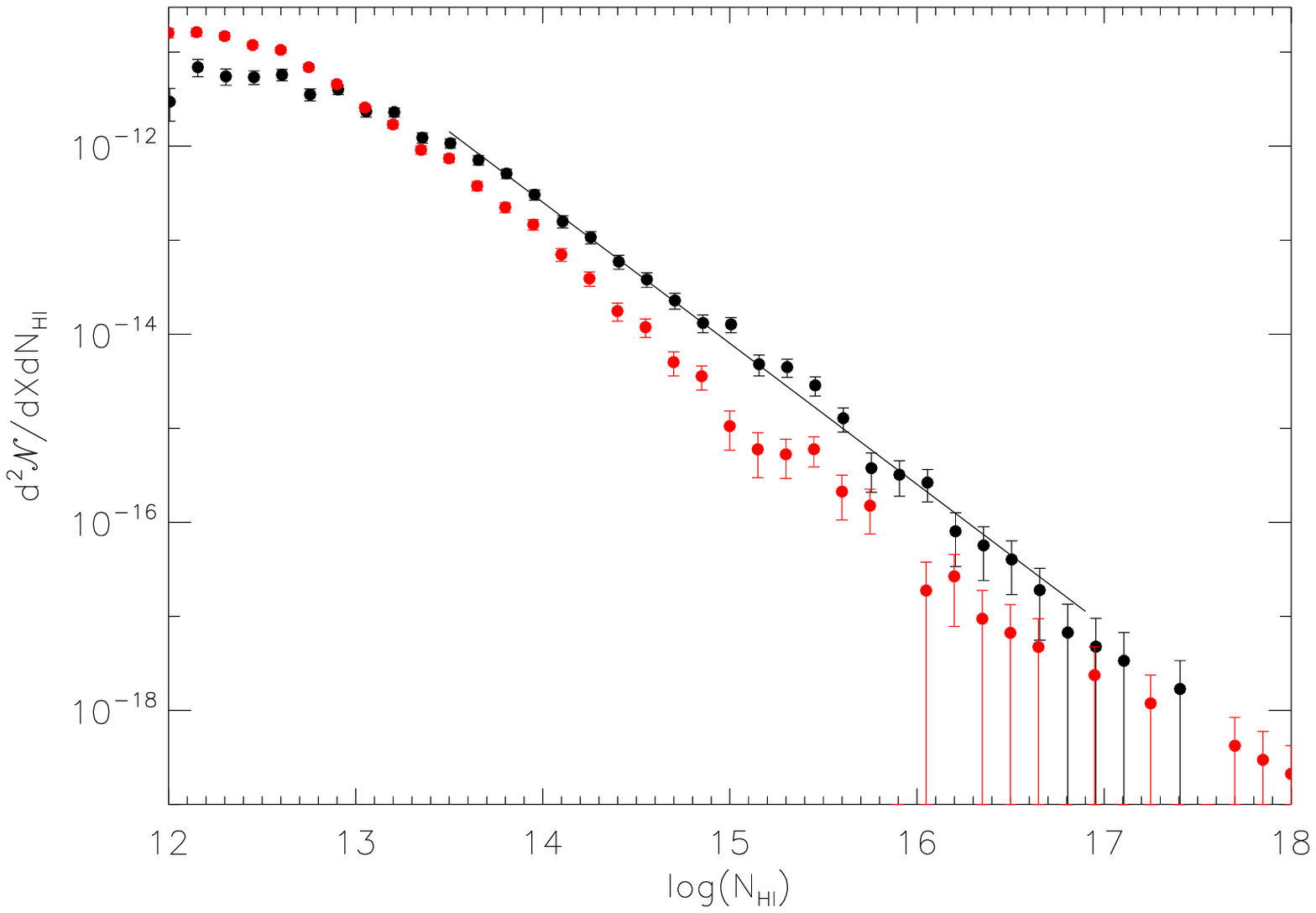}
\caption{The \hi Column Density Distribution Function, $f_{(\nhi,X)}$
  as measured in three sightlines at $z\sim 4.3$ (black points), and
  five sightlines at $z\sim 2.4$ (red points).  The distributions are
  offset by roughly 0.5 dex, or a factor of three in normalization.
  Solid line shows a $N^{-1.5}$ power law, which was used as input for
  the Monte Carlo calculations described below.}
\label{fig:cddf}
\end{figure}

In total, our \hi sample contains 139 lines, with 13 additional
candidates discarded from the sample because of saturation in the
Lyman series.  Just one of the discarded systems is associated with a
possible \civ absorber.

When using high order Lyman transitions to constrain $\nhi$, there is
danger of over-estimating the column because of blended \lya
absorption at lower redshift contaminating the signal.  To assess this
potential error, we generated Monte Carlo realizations of the \lya
forest in our redshift range to test whether the fitting procedure
introduces systematic errors.

The details of \lya forest line samples are not extensively
characterized above $z\sim 3.5$ and it is beyond our present goals to
provide a comprehensive statistical analysis of the \hi forest.
However, as a byproduct of our fitting procedure, we obtain a
determination of the \hi column density distribution at $z\sim 4.2$,
shown in Figure \ref{fig:cddf}.  As at lower redshift, it follows a
power-law form with incompleteness roll off at
$\nhi<10^{14}$\pcmsq---well below the cutoff $\nhi$ of our \civ
sample.  \citet{kim_forest} analyzed \lya forest spectra at $z=2.31$
and $z=3.35$, finding $f(N_\mhi)\propto \nhi^{-1.46}$.  This same
shape provides an accurate fit to our observed $f(N)$ if a slightly
higher overall normalization is used to account for the higher density
of forest lines at $z\sim 4.2$.

For the Monte Carlo tests, we {\em assumed} this form for $f(N)$, with
an overall normalization evolving as $(1+z)^{2.17}$ \citep{kim_forest}
to approximate the evolution in density of lines.  Artificial spectra
were generated using lines drawn from these distributions, with the
full Lyman series included for each absorber.  \lya lines at lower
redshift were intentionally added into the \lyb and higher order
regions at appropriate density to simulate blending and contamination.
Finally, Gaussian noise was injected into each spectrum with an
amplitude set by the noise arrays in our sample spectra.  We do not
claim the resulting spectra to be a perfect representation of the \lya
forest at $z>4$, but they form a useful tool for bootstrapping
estimates of our systematic errors in $\nhi$ measurement from blending.

\begin{figure}
\epsscale{1.0}
\plotone{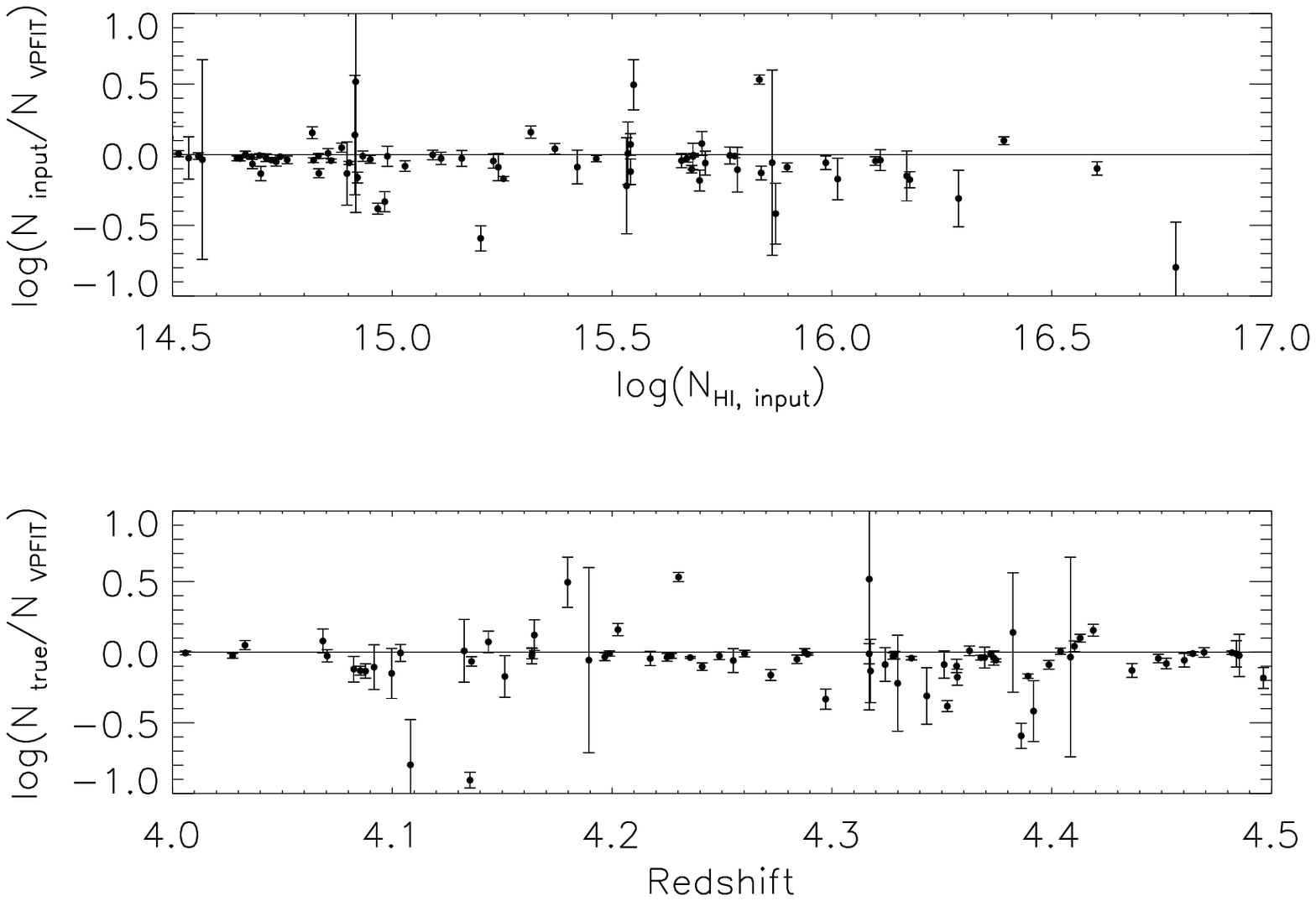}
\caption{Input vs. recovered column densities in our Monte Carlo tests
  of the {\tt vpfit} method for selecting the \hi sample.  We detect
  an overall scatter of $\sim 0.1$ dex in determining $\nhi$, with
  systematic offset of 0.04 dex.  Ionization corrections render the
  effective error in [C/H] smaller than the error in $\nhi$ by a
  factor of 3.}
\label{fig:montecarlo}
\end{figure}

We ran VPFIT on our simulated spectra without knowledge of the input
parameters, in identical fashion to the true data set.  Figure
\ref{fig:montecarlo} shows a comparison of the best-fit $\nhi$
determinations with the true input values, as a function of $\nhi$ and
redshift.  We find a scatter of $\pm 0.1$ dex between the true and
fitted values, with a small systematic (median) offset of 0.04 dex in
the sense that the fitted $\nhi$ over-estimate the true values.
Individual outliers can miss by 0.5-0.7 dex in either direction, but
[65,85]\% of systems are determined to better than [0.1,0.2] dex.  

\subsection{Comparison Sample}\label{sec:z2.5sample}

To study evolutionary trends, we compare the $z\sim 4.3$ \civ sample
with a separate sample of absorbers at $z\sim 2.4$, taken from
previously published work \citep{simcoe2004}.  In this work, \ovi
column densities were measured for a sample of 230 absorbers with
$\nhi \ge 10^{13.6}$\pcmsq.  A smaller sample of 83 systems were used
to measure \civ absorption at $\nhi \ge 10^{14}$ \pcmsq.

\begin{figure}
\epsscale{1.0}
\plotone{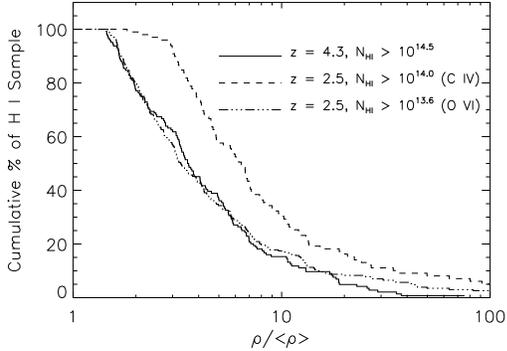}
\caption{Distribution of overdensity in our \hi-selected samples at
  $z\sim 4.3$ and $z\sim 2.4$.  The \civ sample at $z\sim 4.3$ probes
  smaller overdensities than the $z\sim 2.4$ \civ sample, by a factor
  of 2.  The distribution is very similar to \ovi samples at lower
  redshift.}
\label{fig:km_oden}
\end{figure}

In Figure \ref{fig:km_oden}, we plot the cumulative distribution of
$\oden$ for each sample.  We estimate the overdensity at each \hi
absorber location using the formalism of \citet{schaye_civ_pixels}:
\begin{equation}
n_H = 10^{-5}~{\rm cm^{-3}}\left({{\nhi}\over{2.3\times 10^{13}{\rm ~cm^{-2}}}}\right)^{\frac{2}{3}} T_4^{0.17}\Gamma_{12}^{\frac{2}{3}}\left({{f_g}\over{0.16}}\right)^{-\frac{1}{3}}.
\end{equation}
Here, $T_4$ represents the temperature in units of $10^4$K,
$\Gamma_{12}$ represents the flux of ionizing photons at the \hi
ionization edge, and $f_g$ represents the gas to total mass fraction
in \lya forest systems, which is assumed to be near
$\Omega_b/\Omega_M$.  To obtain overdensity, we normalize this density
by the mean gas density, $\Omega_b\rho_c(1+z)^3$.  We assume $T_4$ to
be constant and unity over the $z=2.4-4.3$ range \citep{schaye_temp}.

Figure \ref{fig:km_oden} shows that the \civ sample at $z=4.3$ probes
a median {\em overdensity} that is $2\times$ lower than the \civ
sample at $z\sim 2.4$, despite having a limiting $\nhi$ that is
$3\times$ higher.  The \ovi sample at lower redshift probes almost an
identical range in gas overdensity to the high redshift \civ sample.
Therefore, comparing \civ at both redshifts removes any ambiguities in
the [C/O] relative abundance but means that the high redshift sample
is probing lower densities.  On the other hand, comparing [O/H] at
$z\sim 2.4$ and [C/H] at $z\sim 4.3$ traces the abundance evolution at
fixed overdensity for an assumed value of [O/C].

 Unfortunately this quantity is degenerate with one's choice of UV
 background spectrum used for calculating ionization corrections
 \citep{simcoe2004, schaye_civ_pixels, aguirre_ovi_pixels}.  In the
 discussions below we assume [O/C]$>0$ in the IGM, consistent with
 ionization by a soft spectrum composed of both galaxies and quasars
 \citep{haardt_cuba}.  This model is favored by pixel optical depth
 measurements of \ciii/\civ and \ovi/\siiv \citep{schaye_civ_pixels,
   aguirre_ovi_pixels}.

We caution however that observations of the most metal-poor damped
\lya systems show a recovery toward [O/C]=0 at very low
metallicities\citep{pettini_oc}.  Solar [O/C] ratios can be achieved
with a hard background composed of quasar light with negligible galaxy
contribution.  

\subsection{\civ Sample Definition and Measurements}\label{sec:selection}

For each of the \hi absorbers in the sample, we examined the
corresponding wavelength of the \civ doublet.  Using VPFIT, we fit
\civ absorption profiles to any absorbers within a velocity range of
$\pm 20$\kms ~of the \hi centroid.  In blended systems, we matched
individual \civ components to their corresponding \hi components where
possible.  If a single \hi line contained multiple components of \civ
then the total column density for these components was recorded.
Finally, if no \civ is detected we used VPFIT to calculate $1\sigma$
upper limits on the \civ column density, using the error array.

\begin{figure}
\epsscale{1.0}
\plotone{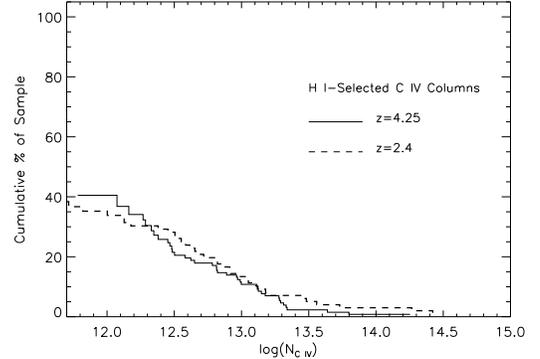}
\caption{Cumulative distribution of \civ column densities at $z\sim
  4.3$ and $z\sim 2.4$.}
\label{fig:civ_km}
\end{figure}

The final sample therefore consists of a mix of detected systems with
measured column densities, and non-detections with upper limits.  We
used a $\ge 3\sigma$ cut as the threshold between detections and
non-detections.  For the non-detections, VPFIT usually finds a formal
solution $\nciv\pm\sigma_{\mciv}$ with $\nciv \le 3\sigma_{\mciv}$;
the upper limit is then $\nciv + 3\sigma_{\mciv}$.

The final sample of $z=4.0-4.5$ \hi and \civ redshifts and column
densities is presented in Table 2.  The three sightlines contain 131
systems with $\nhi > 10^{14.5}$ Among this sample, 39 lines have \civ
detections, while 92 are non-detections.  The approximate limiting
\civ column density is between $1-2\times 10^{12}$\pcmsq.

\section{Cumulative Distributions of \civ, and $\nciv/\nhi$}\label{sec:km_lines}

Figure \ref{fig:civ_km} shows the distribution of \civ column
densities in samples drawn at the two different redshifts considered.
Because the \civ samples contain a mixture of detections and upper
limits, we have plotted the cumulative distribution as determined by
the Kaplan-Meier estimator.  The Kaplan-Meier product limit is a
fundamental tool of survival statistics, which deals with censored
datasets such as the one presented here \citep{feigelson_survival}.
Its application for measuring abundances from absorption lines is
discussed in \citet{simcoe2004}.

For the Kaplan-Meier distribution to accurately represent the true
distribution, two conditions must be satisfied.  First, the upper
limits must be statistically independent of one another, which is
clearly true for the resolved, discrete absorption lines in our
sample.

Second, the probability of a non-detection must not depend on the
criterion used to select the sample.  Since we are measuring \civ
detections and upper limits for a sample selected on the basis of \hi
column density, this condition should hold.  There may be slight
discrepancies at the very high end of the \hi sample because of local
chemical enrichment.  But because the \hi column density distribution
is a steep power law, the vast majority of our sample is at the low
end where the censoring only depends on variation in signal-to-noise
in the \civ region and is therefore quite random.

Returning to Figure \ref{fig:civ_km}, it is clear that the
distribution of \civ evolves very little between $z\sim 2.4$ and
$z\sim 4.3$.  The fact that the cumulative distributions only reach
$\sim 40\%$ stems from the survival analysis --- recall that \civ is
only detected in 39 of 131 systems, or 30\%.  Most of the upper limits
are clustered near low $\nciv$, but some are at slightly higher
values.  Kaplan-Meier accounts for the difference between the fraction
of detections (30\%) and the value of the cumulative distribution near
the limiting $\nciv$ (40\%).

\begin{figure}
\epsscale{1.0}
%\plotone{/Users/Rob/z4civ/plots/civ_over_hi.ps}
\plotone{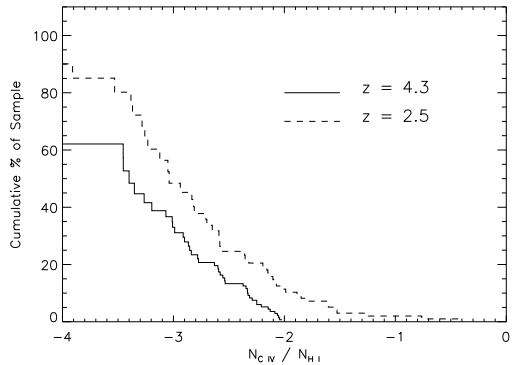}
\caption{Cumulative distribution of the \civ/\hi ratio at $z\sim 4.3$
  and $z\sim 2.4$.  This is the fundamental observed quantity in our
  analysis. }
\label{fig:civhi_km}
\end{figure}

The constancy of the \civ distribution with redshift has been noted by
other authors, most notably \citet{songaila_new_civ}.  Our result is
consistent with these previous findings.

For studies of the IGM metallicity, our fundamental observable is
$\nciv/\nhi$.  Because \civ evolves very little with redshift but the
\hi opacity increases, we expect the \civ/\hi ratio to decrease toward
higher redshift.  The cumulative distribution of this quantity is
shown in Figure \ref{fig:civhi_km}, with the expected difference
between $z\sim 2.4$ and 4.3.  Note that this is the distribution of
the $\nciv/\nhi$ ratio for each system in the sample, not the ratio
of the \civ and \hi distributions.  The shapes are very similar at the
two epochs, with the difference coming from a systematic shift
downward by roughly 0.4 dex, or a factor of 2.5.

\section{Ionization Corrections}\label{sec:ionization_corrections}

To translate the evolution in $\nciv/\nhi$ into an evolution in the
carbon abundance, we apply an ionization correction to each individual
absorber as follows:
\begin{equation}
\left[{{C}\over{H}}\right] = \log\left({{{\nciv}\over{\nhi}}}\right) +
\log\left({{f_{\mhi}}\over{f_\mciv}}\right) - 
\log\left({{C}\over{H}}\right)_\sun.
\label{eqn:metallicity}
\end{equation}
Here, $f_\mciv$ and $f_\mhi$ represent the triply ionized fraction of
carbon, and neutral fraction of hydrogen, respectively.  The last term
is a zero-point offset to normalize relative to the solar abundance.
Throughout the paper we have assumed the meteoric Solar abundances of
\citet{grevesse_solar_abund}, where $A_{\rm carbon}=8.52$ on a scale
with $A_{\rm Hydrogen}=12$.  Several revisions have been proposed to
the Grevesse solar abundances \citep{allende_prieto_oc,
  allende_prieto_oxygen, holweger_solar_abundances}; we have kept them
for consistency with prior work by many authors on IGM abundances.

Ionization fractions for carbon and hydrogen were determined using the
CLOUDY \citep{cloudy} software package.  We assumed the IGM is
optically thin (consistent with the low \hi column densities in the
sample), and in photoionization equilibrium.

\subsection{Comments on choice of the UV background spectrum}\label{sec:uvbg}

\begin{figure}
\epsscale{1.0}
\plotone{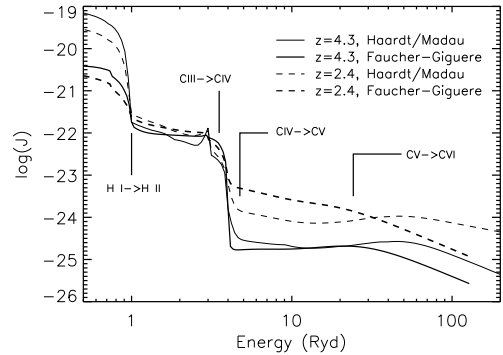}
\caption{Mean spectrum of the UV background radiation field, at $z\sim
  4.3$ and $z\sim 2.4$.  We show the calculations of both
  \citet[including a galaxy contribution]{haardt_cuba} and
  \citet{faucher_uvbg} for comparison.  The spectra are very similar
  at high redshift; at lower redshift the \citet{faucher_uvbg} version
  is harder, leading to higher overall abundances and a
  correspondingly larger evolution in the observed [C/H] ratio.}
\label{fig:uvbg}
\end{figure}

The largest uncertainty in our ionization correction arises from the
choice of a particular prescription for the ionizing background
spectrum.

Several groups have recently provided improved constraints on the
overall intensity of the background field at 1 Rydberg - denoted as
$J_{912}$.  This can be derived from the opacity of the \lya forest,
which relates in turn to the \hi photoionization rate $\Gamma_{12}$.
In a sample of 63 high resolution spectra spanning $2.0 < z < 5.5$,
\citet{becker_gamma} find $\Gamma_{12}=[0.8, 0.5]$ at $z=[2.4, 4.3]$,
where by convention $\Gamma_{12}$ is normalized to units of
$10^{-12}{\rm s}^{-1}$.  \citet{faucher_gamma} find $\Gamma_{12}=[0.6,
  0.4]$ over the same range using slightly different analysis
techniques on a set of 86 high resolution spectra.  Earlier analyses
by \citet{bolton_gamma} and \citet{scott_uv_bkgd} derived slightly
higher estimates closer to $\Gamma_{12}=1$.  However, all of these
studies find that the evolution of the \hi photoionization rate is
quite weak over the redshift range studied here, perhaps $10-30\%$
smaller at $z=4.3$ but not much more.  Throughout the rest of the
paper, we use the values from \citet{becker_gamma}.  For a power law
ionizing spectrum $\propto \nu^{-1.8}$, this corresponds to a flux at
the \hi ionization edge of $\log(J_{912})=[-21.5,-21.7]$ at $z=[2.4,
  4.3]$.

Even if the normalization of the background spectrum at 1 Rydberg does
not evolve, it is generally agreed that the {\em shape} of the
spectrum will change.  The quasar luminosity function falls off
between $z=4$ and $z=2$ \citep{hopkins_lf}.  While ionizing photons at
$z=2.4$ come from both quasars and galaxies, the background at $z=4.3$
contains a higher fractional contribution from galaxies
\citep{faucher_gamma}.  This should lead to a softer UV background at
earlier epochs, with fewer photons at the higher energies which
regulate the ionization balance of heavy elements like carbon and
oxygen.

\begin{figure}
\epsscale{1.0}
\plotone{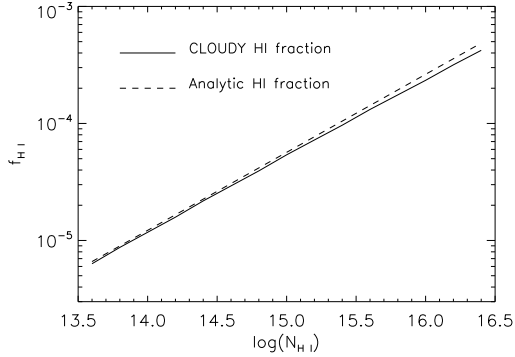}
\caption{Neutral fraction of hydrogen as calculated by CLOUDY,
  compared with the analytic approximation of Equation 3.  This
  relation is redshift-independent, except for the implicit dependence
  on the IGM temperature and $\Gamma_{12}$.  It has a logarithmic
  slope of $\frac{2}{3}$ dex change in $f_\mhi$ per unit dex in
  $\nhi$.}
\label{fig:fhi}
\end{figure}

We have tested two main prescriptions of the UV background with our
data: the commonly used version from \citet{haardt_cuba}, and an
independent new determination from \citet{faucher_uvbg}.  In Figure
\ref{fig:uvbg} we show these spectra, with the energies of relevant
transitions for the \civ ionization balance labeled.  
The Haardt~ \& Madau spectrum includes contributions from both QSOs and
galaxies; the QSO portion is determined from the \citet{croom_qso_lf}
luminosity function, while the galaxies are added through population
synthesis modeling, assuming a 10\% escape fraction of photons beyond
the Lyman limit.  The Haardt\& Madau spectra have been normalized to
match the observational constraints on $\Gamma_{12}$ described above.
\citet{faucher_uvbg} take the QSO luminosity function from
\citet{hopkins_lf} which falls more steeply toward higher redshift,
and add a galaxy contribution to keep the evolution of $\Gamma_{12}$
constant.

When the Haardt \& Madau spectrum is renormalized in CLOUDY to match
observations of $\Gamma_{12}$ at 13.6 eV, its soft X-ray background is
also artificially boosted by a factor of $\sim 10$ at energies above 10
Rydbergs.  This has consequences for the \civ ionization balance
because it affects the \cv to \cvi transition.

\begin{figure}
\epsscale{0.95}
\plotone{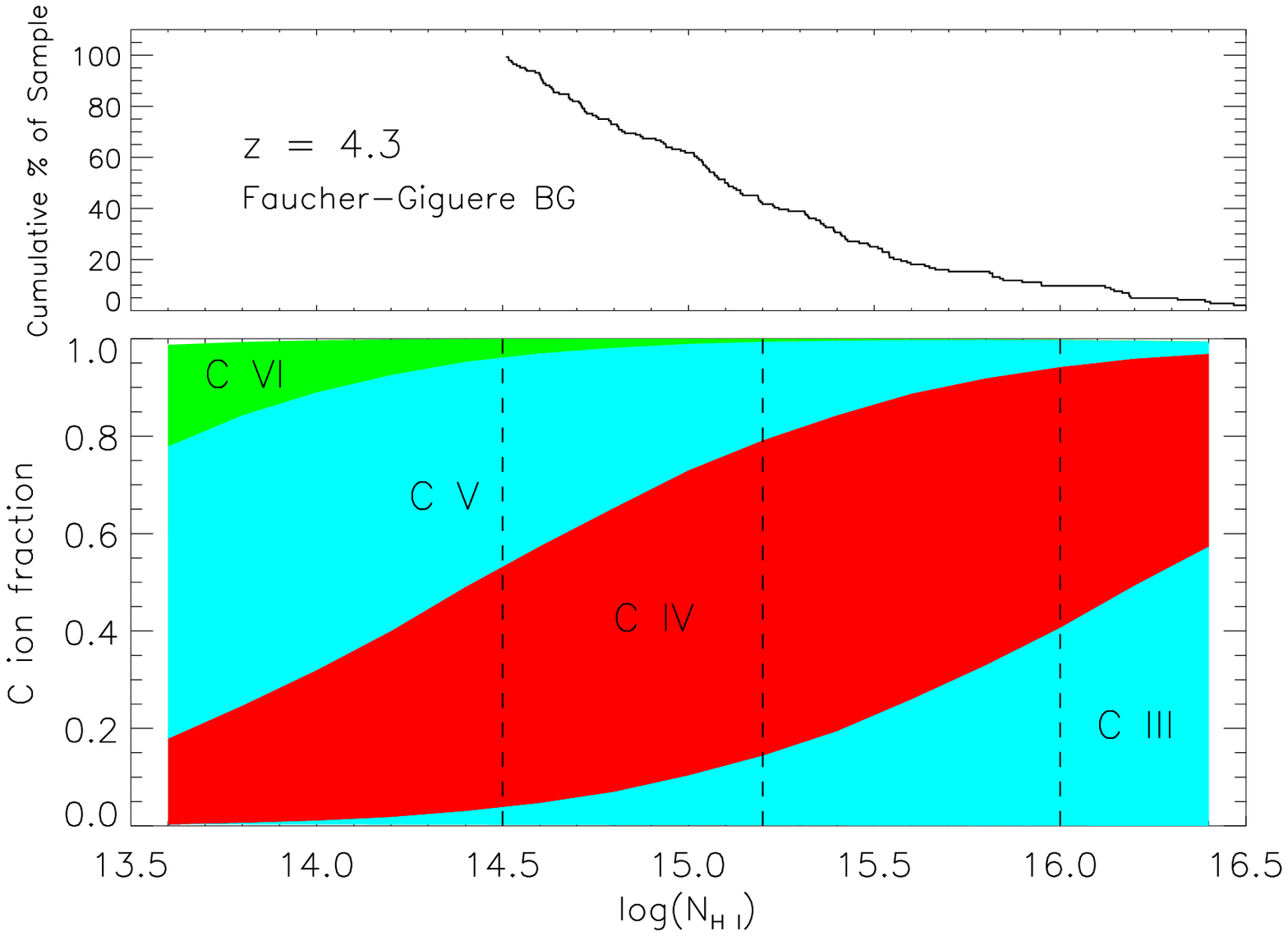}
\plotone{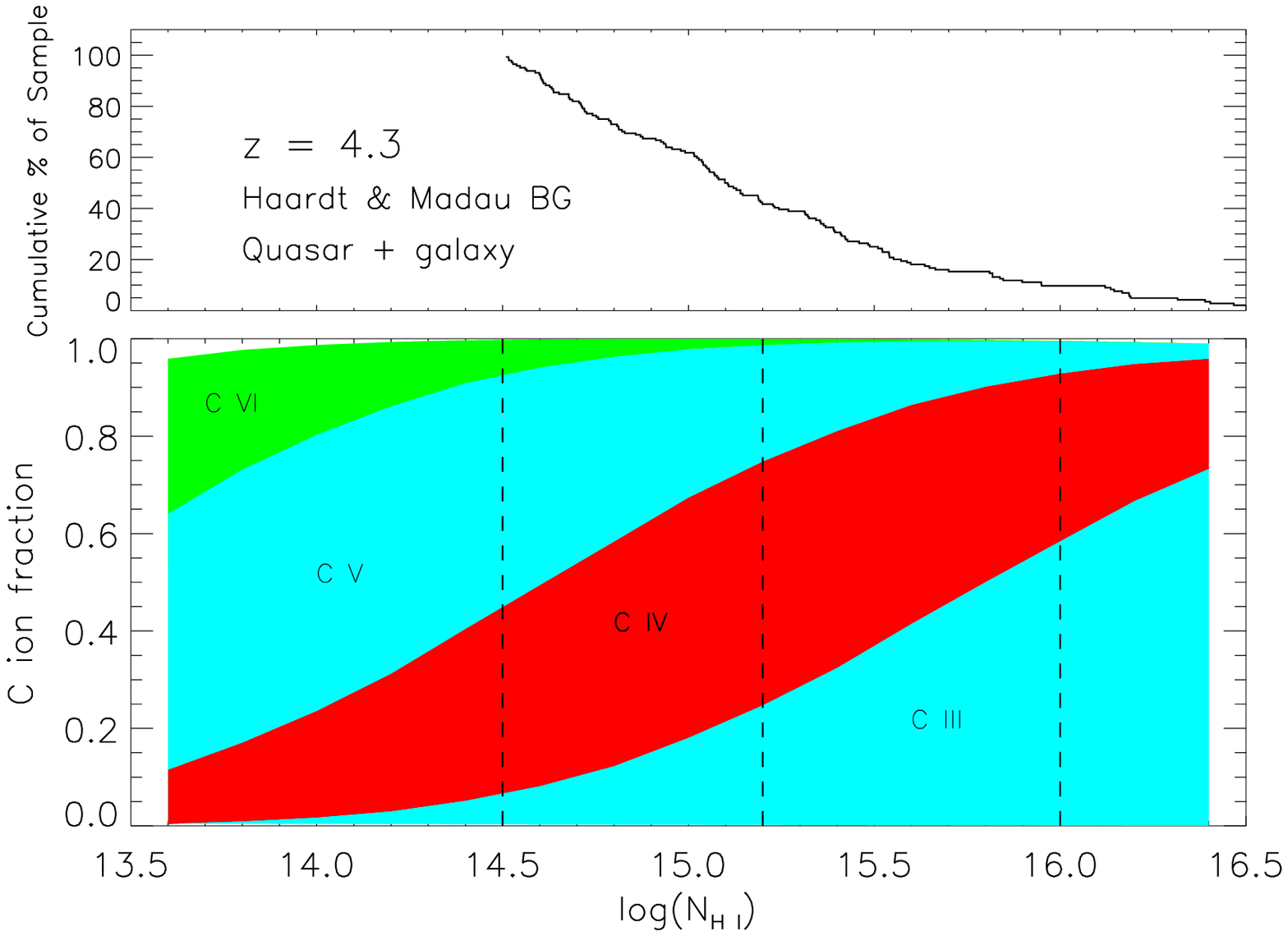}
\caption{CLOUDY carbon ionization balance at redshift $4.3$ for the
  \citet{haardt_cuba} model of the UV background (top) and
  \citet[bottom]{faucher_uvbg} models.  The histogram at top
  indicates the cumulative distribution of lines in our sample, with
  the minimum, median, and 90th percentile shown by vertical lines.
  At this redshift, the \civ fraction peaks in the IGM, at
  $f_\mciv\sim 0.5$.}
\label{fig:fciv_z4}
\end{figure}

In their calculations, Haardt \& Madau treat the UV and X-ray
backgrounds independently, trying to match each with observed
luminosity functions by balancing the relative importance of Type I
and II QSOs.  In its raw form, the Haardt \& Madau X-ray background is
slightly lower at z=4.3 than at 2.4 (as would be expected), but after
renormalization to matched the observed $\Gamma_{12}$ at 1 Ry, the
situation reverses, such that the X-ray background is higher at
$z=4.3$ than it is at $z=2.4$.  Since the X-rays originate in AGN
which are proportionately fewer at high redshift, this situation is
probably unphysical.  So, we applied a downward correction of 0.8 dex
to the $z=4.3$ HM spectrum above 10 Ryd to soften it by a comparable
amount as the UV background and bring it back into agreement with the
original X-ray intensity.  Coincidentally, this downward correction
brings the HM spectrum into fairly close agreement with the
\citet{faucher_uvbg} spectrum, which by construction does not suffer
from this effect of renormalization.

While the HM and \citet{faucher_uvbg} spectra do not agree in every
detail, their general shapes match fairly well, particularly at
$z=4.3$.
At lower redshift the difference is larger owing to assumptions on the
AGN contribution as well as the treatment of \heii reionization.
However, in both models the main change from $z=2.4$ to $z=4.3$ is a
factor of 10-20 decrease in the hard UV background.  It should be
noted that neither model captures the \heii ``sawtooth'' effect
described in \citet{HM_sawtooth}; this is discussed in Section
\ref{sec:sawtooth}.

\subsection{\hi ionization fractions}\label{sec:fhi}

Since the \lya forest is optically thin and in photoionization
equilibrium, a calculation of $f_{\mhi}$ is relatively straightforward
and may be made even without detailed knowledge of the spectral
shape.  In equilibrium, the ionization and recombination rates may be
balanced as:
\begin{equation}
f_{\mhi}n_{H} \Gamma \approx n_H^2 R
\end{equation}
where the recombination rate $R\approx 4\times 10^{-13}T_4^{-0.76}$
cm$^{3}$s$^{-1}$, $n_H$ is given in Equation 1, and we assume the gas
is mostly ionized.  Equations 1 and 3 may be combined to derive the
hydrogen ionization fraction as a function of (observed) \hi column
density, plotted in Figure \ref{fig:fhi} alongside its numerical
CLOUDY determination.  This relation is independent of redshift,
except implicitly through the (weak) evolution of the ionization rate
$\Gamma$, and the temperature $T_4$.

\subsection{\civ ionization fractions}\label{sec:fciv}

\begin{figure}
\epsscale{0.95}
\plotone{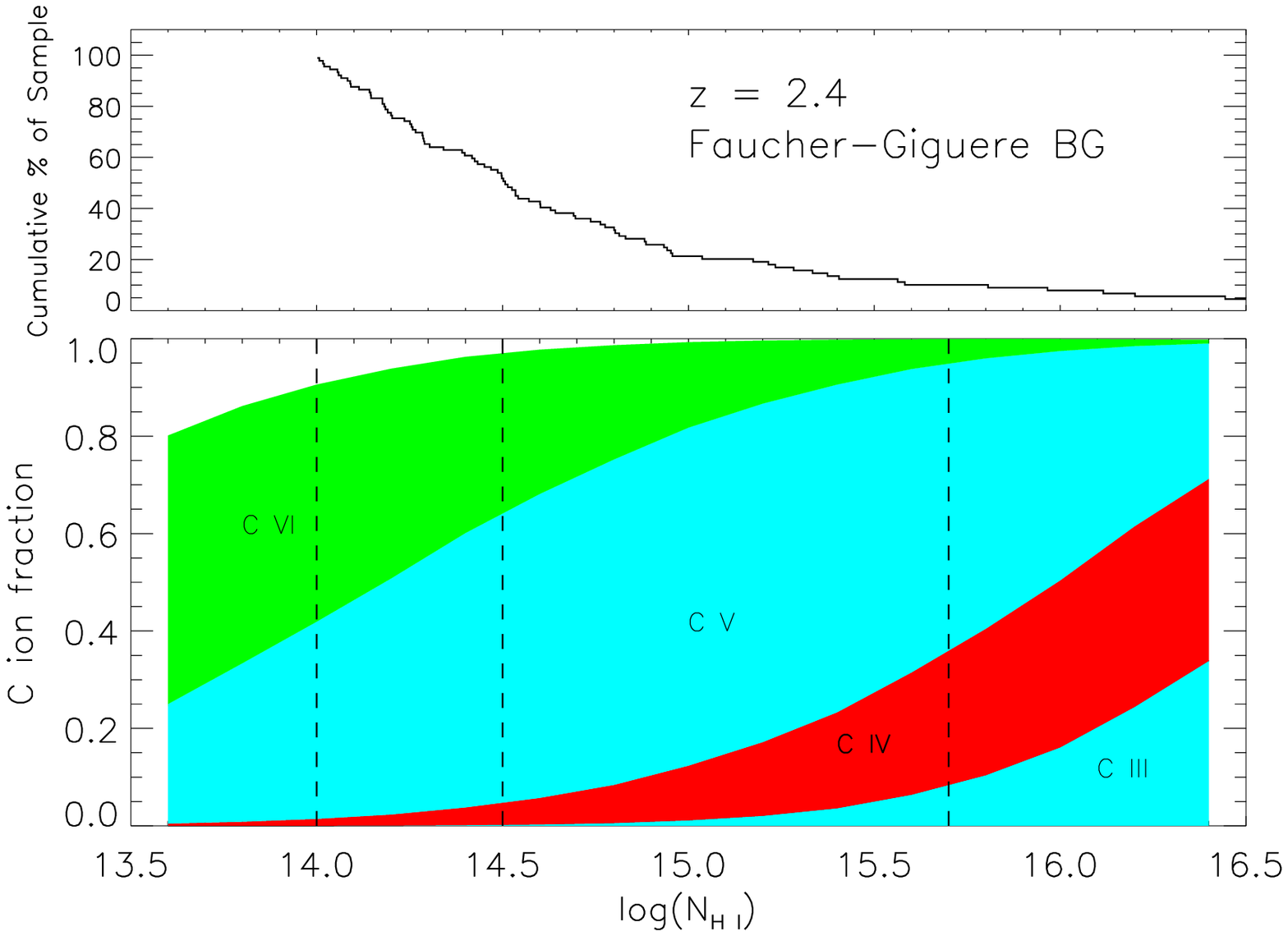}
\plotone{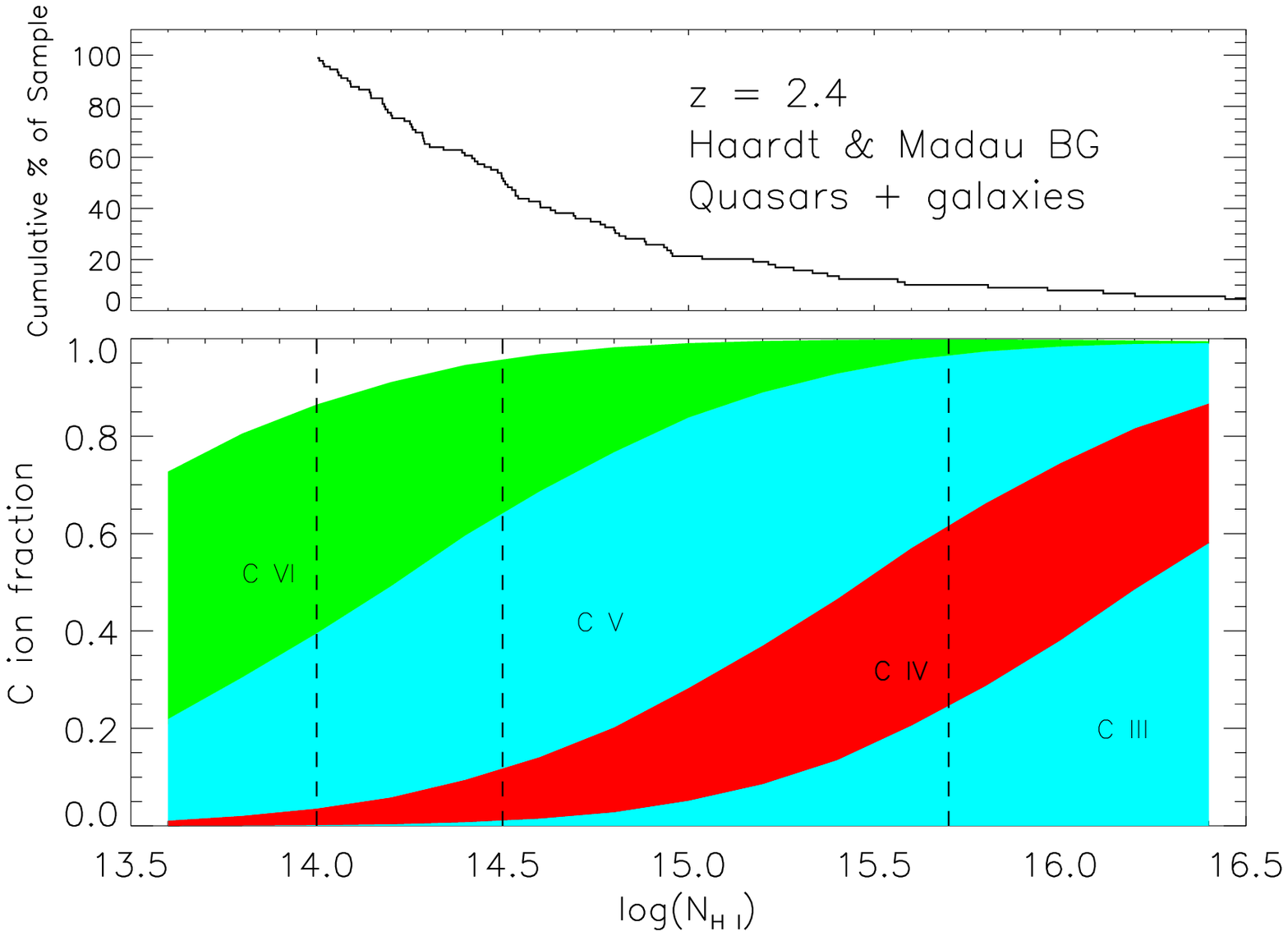}
\caption{CLOUDY carbon ionization balance at $z\sim 2.4$, again for
  both models of the UV background.  The mean \hi column density is
  lower, as is the ionization fraction in \civnsp.  The lower
  ionization fraction may be directly mapped to the inferred change in
  abundance.}
\label{fig:fciv_z2}
\end{figure}

The ionization balance of \civ is a far more difficult calculation,
because it is governed by photons at frequencies where the UV
background spectrum is less well constrained observationally.  Figure
\ref{fig:fciv_z4} shows the CLOUDY-derived ionization balance for
carbon as a function of \hi column density, at $z=4.3$ (the balance at
$z=2.4$ is shown in Figure \ref{fig:fciv_z2}).  To calculate the
ionization fractions, we first estimated the total hydrogen density
associated with each \hi column density as calculated from Equation 1.
This value of $n_H$ was input into CLOUDY along with the prescriptions
shown in Figure \ref{fig:uvbg} for the shape and amplitude of the
background spectrum.  Together, the determinations of the density and
ionizing background flux determine the ionization parameter $U$ for
the cloud which allows CLOUDY to calculate the abundance fraction of
each carbon ionization state.

The top panel of each figure shows the cumulative distribution of \hi
column densities for the \civ samples, as a visual aid to indicate
what ionization corrections are appropriate for which fractions of the
sample.  Vertical lines on the bottom panels indicate the sample
minimum $\nhi$, and also its median value and 90th percentile.  Plots
are shown both for the Haardt \& Madau and Faucher-Giguere forms of
the background spectrum.

These plots show that the \civ ionization state predominates at $z\sim
4.3$, at a value of $f_\mciv = 0.50-0.65$ for the median \hi system in
the sample.  This range reflects the difference between the
\citet{haardt_cuba} and \citet{faucher_uvbg} backgrounds, respectively.
Roughly equal amounts of carbon are in higher and lower ionization
states.

Conversely, the median system at $z\sim 2.4$ contains only 4-10\%
\civ, with the vast majority of carbon inhabiting higher ionization
states.  So, the typical ionization correction at $z=4.3$ is an
upwards factor of 1.5-2.0 (0.2-0.3 dex), compared to a correction
factor of $10-25$ (1.0-1.4 dex) at lower redshift.

\section{Abundance Results}\label{sec:abundance}

\subsection{Evolution from $z=4.3$ to $z=2.4$: Qualitative Results}\label{sec:qualitative}

Before presenting the full carbon abundance distribution results, it
is instructive to examine the evolution of the median system using
qualitative arguments.  From Figure \ref{fig:civhi_km}, the median \hi
absorber decreases in its \civnsp:\hi ratio by 0.4 dex between $z=4.3$
and $z=2.4$.

Suppose now that there were no evolution in the [C/H] ratio. To
achieve this, our measured 0.4 dex decrease in \civ/\hi would need to
be offset by a corresponding 0.4 dex {\em increase} in
$\log\left({f_\mhi}/{f_\mciv}\right)$, according to Equation 2.

We have seen already that $f_\mhi$ depends on $\nhi$ but only very
weakly on redshift.  For a median $\log(\nhi)=14.5$ at $z=2.4$ and
$15.2$ at $z=4.3$, the \hi ionization fractions can be read from
Figure \ref{fig:fhi}, yielding $\log (f_{\mhi,4.3})-\log
(f_{\mhi,2.4}) \approx 0.5$.

Thus for fixed $f_\mciv$, the first term in Equation 2 is 0.4 dex
{\em lower} at $z\sim 4.3$, while the second is 0.5 dex {\em higher}.
The first term is observationally determined, and the second is fairly
model-independent, and they approximately cancel each other out.  The
degree of evolution in [C/H] therefore corresponds almost exactly to
the change in $f_\mciv$ with redshift.

Even if we take the most conservative possible hypothesis - that the
UV background spectrum at $z=4.3$ is identical to the spectrum at
$z=2.4$ - then $U$ will decrease because of the increase in gas
density at higher redshift.  But observational studies of the QSO and
galaxy luminosity functions at these redshifts indicate that the
background spectrum should soften as it is increasingly dominated by
galaxies toward higher redshift \citep{faucher_gamma}.  These factors
cause an increase in $f_\mciv$ which leads to a decrease in implied
[C/H].  Thus the observed distribution of $N_\mciv/N_\mhi$, coupled
with minimal assumptions about ionization, suggest that [C/H] is lower
at $z=4.3$ than it is at $z=2.4$.

To estimate by how much, consider the CLOUDY calculations from both
the \citet{haardt_cuba} and \citet{faucher_uvbg} spectra.  These
illustrate that the \civ ionization fraction for the median system
increases by 0.7 dex toward higher redshift.  Taken together, these
qualitative arguments suggest that [C/H] in the median system is lower
at $z\sim 4.3$ by 0.5-0.6 dex (a factor of $\sim 3$) than it is at
$z=2.4$.  

They also suggest that the {\em scatter} in the \civ to \hi ratio may
be an interesting diagnostic of fluctuations in the hard UV
background.  The scatter in this ratio reflects the convolved scatter
in the intrinsic [C/H] ratio, measurement errors, and variance in the
ionization correction.  Even if one makes no assumptions about the
relative weighting of these three factors, one may set upper bounds on
the scatter in the hard background under the conservative assumption
of zero measurement error and abundance scatter.

\subsection{The Cumulative [C/H] Distribution at $z\sim 4.3$}\label{sec:km_metals}

\begin{figure}
\epsscale{1.0}
\plotone{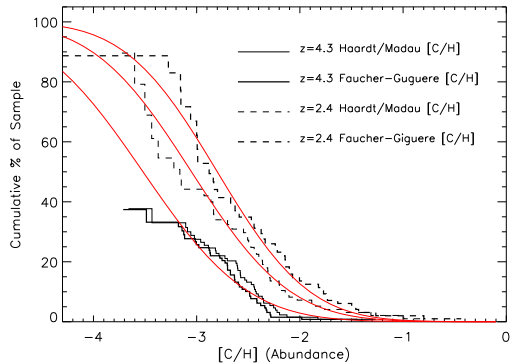}
\caption{Kaplan-Meier cumulative probability function for carbon
  abundance in the IGM at $z\sim 4.3$ and $z\sim 2.4$.  Note that the
  low redshift curves sample the IGM at a median overdensity roughly
  $2\times$ larger than the high redshift sample.  Solid curves
  indicate lognormal probability distributions.  At $z\sim 4.3$, the
  curve represents a median abundance of [C/H]=-3.65, and
  $\sigma=0.8$.  At lower redshift the median rises to [C/H]=-3.1 for
  the \citet{haardt_cuba} background, or [C/H]=-2.8 for
  \citet{faucher_uvbg}, with similar scatter.  The difference of
  0.5-0.7 dex represents a $3-5\times$ increase in the median
  abundance for the low redshift sample.}
\label{fig:km_ch}
\end{figure}

Figure \ref{fig:km_ch} shows the Kaplan-Meier distribution of [C/H] at
$z=4.3$ and $z=2.4$, derived by applying ionization corrections
individually to each \civ system (or its upper limit).  The
distributions are shown for both the \citet{haardt_cuba} and
\citet{faucher_uvbg} forms of the UV background radiation field.
Recall that we show the [C/H] distributions for each redshift to
eliminate ambiguity in [O/C], but the $z=4.3$ sample probes gas with a
median overdensity of $\oden \approx 3$, whereas the \civ sample at
$z=2.4$ only reaches a median $\oden\approx 6$ (the sample minima are
$\oden\approx 1.6$ and $\oden \approx 3$).

At $z=4.3$, our sample reaches nearly to the median [C/H] ($50\%$ on
the plot), which lies near [C/H]$=-3.55$, or $\sim 1/3500$ the Solar
abundance.  Roughly half of the lines in our sample have $3\sigma$
upper limits restricting them below this value, so the shape of the
distribution below the median is not known.  Above the median, the
distribution appears to be very crudely lognormal, with the upper
quartile falling above [C/H]$=-3.1$.  The distributions are very
similar for the two choices of UV background radiation spectrum at
$z=4.3$.  The smooth solid curve shows the cumulative distribution for
a lognormal probability distribution.  Our best fit lognormal model
shows a median abundance of [C/H]=-3.55, and lognormal deviation of
$\sigma=0.8$ dex.

At $z=2.4$, we find a median abundance of [C/H]$=-3.1$ with
$\sigma=0.8$ dex for the \citet{haardt_cuba} form of the background
spectrum, consistent with prior results of \citet{simcoe2004} and
close to, though slightly higher than those of
\citet{schaye_civ_pixels}.  The \citet{faucher_uvbg} form of the
background spectrum results in a slightly higher carbon abundance but
similar scatter, at median [C/H]=$-2.8$ and $\sigma=0.7$ dex.  The
higher median results from the increased flux in the
\citet{faucher_uvbg} spectrum near 5 Rydbergs, at the \civ to \cv
ionization edge.  This flux enhancement favors higher
ionization and hence larger ionization corrections.

Comparing the two distributions, we see that the sample median carbon
abundance has increased by $\sim 0.45-0.65$ dex (depending on the
background model) during the epoch between $z=4.3$ and $z=2.4$,
although the density probed at $z\sim 2.4$ is a factor of 2 larger.
It is interesting to compare this result to that of
\citet{schaye_civ_pixels}, who find $d/dz([{\rm
    C/H}])=+0.08^{+0.09}_{-0.10}$ and $d/d(\log(\rho))({\rm
  [C/H]})=0.65^{+0.10}_{-0.14}$.  Not accounting for density, our
measurement implies a derivative of roughly $-0.26$ dex per unit
redshift, a factor of $3.3$ higher than reported by Schaye et al.
Moreover the evolutionary trend has the opposite sign as Schaye's; our
new result evolves in the expected direction (i.e. increasing
abundance with time), although the prior result was basically
consistent with zero evolution.

Since Schaye et al. reported weak correlation with redshift but strong
correlation with density, it is worth considering whether our measured
evolution is simply an artifact of the different densities probed at
the different redshifts.  According to the regression analysis of
Schaye et al., a $2\times$ change in density alone could account for
$\approx 0.3$ dex difference in [C/H] between the samples, but their
reported evolution of +0.08 dex per unit redshift for $\Delta z=1.9$
would counteract the density effect by 0.15 dex, yielding a total
change in [C/H] of 0.15 dex to our 0.5.  Schaye et al interpret the
small statistical significance of their evolution coefficient and its
non-intuitive sign as a non-detection.  We have detected a larger
difference in [C/H] between our high and low redshift samples, in the
expected direction, even accounting for the difference in density.

To decouple the effects of density and true evolution, we ran our
analysis for a subset of our high redshift sample restricted to
$\log(\nhi)\ge 10^{14.7}$, thus ensuring that [C/H] is measured at
fixed overdensity between the two redshifts.  The resulting
Kaplan-Meier distribution is shifted to higher [C/H] by 0.2 dex for
this factor of 2 increase in $\oden$, suggesting that the abundance is
indeed lower in regions of smaller overdensity, but by a smaller
degree than measured by Schaye et al.  If we budget 0.2 dex of the
total 0.5 change seen in Figure \ref{fig:km_ch} to the change in gas
density, the remaining 0.3 dex represents the temporal evolution
signal.

\begin{figure}
\epsscale{1.0}
\plotone{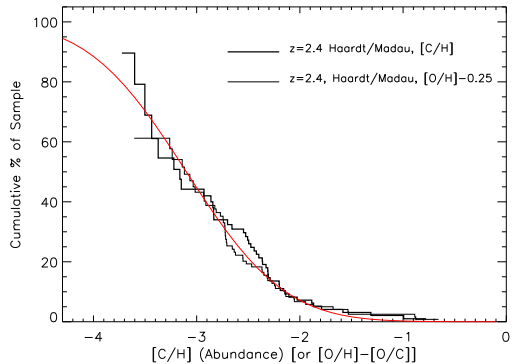}
\caption{Kaplan-Meier Distribution function for Carbon and Oxygen in
  the $z\sim 2.4$ IGM, used for calibrating the [O/C] ratio, which is
  best fit with [O/C]=0.25.  A positive [O/C] has been found by many
  authors, though the value used here is slightly smaller than other
  estimates in the literature (see text).}
\label{fig:km_oc}
\end{figure}

As described earlier, we may also compare our measurements of [C/H] at
$z\sim 4.3$ to pre-existing measurements of [O/H] at $z\sim 2.4$,
which cover an identical range of overdensity bust must be corrected
for any non-Solar value of [O/C].  We estimate the value of [O/C] by
comparing the [O/H] and [C/H] distributions at $z\sim 2.4$, and
applying this same value at $z=4.3$ (i.e. assuming no evolution in
relative solar abundances).  Figure \ref{fig:km_oc} shows the result
of this calculation.  The [C/H] distribution is shown in the thick
solid line, while the [O/H] distribution is the thin line; here the
[O/H] curve has been shifted downward by 0.25 dex, simulating the
predicted [C/H] for a best-fit [O/C]=0.25.  Other studies
\citep{simcoe2004, schaye_civ_pixels, aguirre_ovi_pixels} have
generally found positive [O/C] distributions for soft photoionizing
spectra, with values near [O/C]$\sim 0.5$.  Our slightly smaller value
likely results from a slightly higher normalization of the background
spectrum to match recent observations \citep{becker_gamma,
  bolton_gamma}.  The best-fit [O/C] for the \citep{faucher_uvbg} UV
background is near [O/C]=0, which is to be expected considering that
this spectrum is much harder than the \citep{haardt_cuba} background
at $z=2.4$.

\begin{figure}
\epsscale{1.0}
\plotone{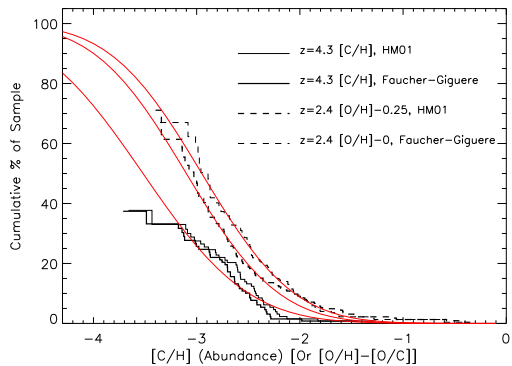}
\caption{Kaplan-Meier Distribution function for [C/H] at $z\sim 4.3$
  and [O/H]-0.25 at $z\sim 2.4$.  Subject to estimates of [O/C], this
  distribution is our best representation of the change in [C/H] at
  fixed overdensity.  The curve at $z\sim 4.3$ is the same as in
  Figure \ref{fig:km_ch}; the lognormal fits shown here for $z\sim
  2.4$ have a median [C/H]=-3.1, -2.9 for the two models of the UV
  background.}
\label{fig:km_constrho}
\end{figure}

Figure \ref{fig:km_constrho} shows the distributions of [C/H] at
$z=4.3$, [O/H]$-0.25$ at $z=2.4$ for the \citet{haardt_cuba} UV
background, and [O/H] for the \citet{faucher_uvbg} model (which has
best-fit [O/C]=0).  Lognormal curves are show for each model.  The two
$z=2.4$ models have median abundances of $-2.9$ and $-3.1$, with
similar scatter.  Evidently there is a 0.5 dex evolution in the median
[C/H] at fixed overdensity for [O/C]=0.25; adopting [O/C]=0.5 would
reduce the absolute change by 0.25 dex.

\begin{figure}
\epsscale{1.0}
\plotone{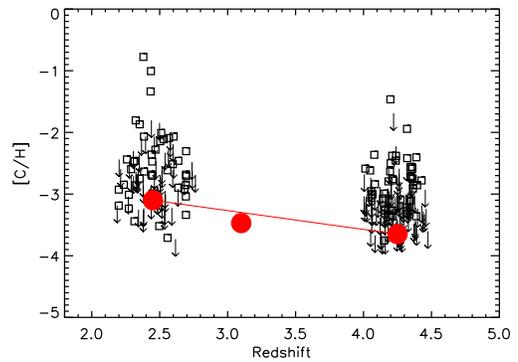}
\caption{Illustration of the individual measurements for each line in
  the full sample.  Large red dots indicate the median abundance at
  each redshift, calculated using the Kaplan-Meier product limit
  estimator.  The median points lie below the center of the scatter
  because of the inclusion of numerous upper limits.  The solid point
  at $z=3.1$ represents the measurement of \citet{schaye_civ_pixels}.}
\label{fig:evolution}
\end{figure}

Figure \ref{fig:evolution} shows this evolution as evidenced in the
individual sample measurements at each redshift.  The large circles
show our median estimates at each redshift, while the central point is
taken from \citet{schaye_civ_pixels}

Taken all together, these calculations indicate that the intergalactic
carbon abundance increased by 0.3-0.5 dex between $z\sim 4.3$ and
$z\sim 2.4$, with little change in the lognormal abundance scatter.
The range quoted reflects uncertainty in various choices for the UV
background, the relation between density and metallicity, and the value
of [O/C] used.  

A change of 0.3-0.5 dex represents a factor of $\sim 2-3$ increase in
the median intergalactic carbon abundance at fixed overdensity.  It
suggests that roughly half of the carbon seen in the IGM at $z\sim
2.4$ was deposited in the 1.3 Gyr interval between $z\sim 4.3$ and
$z\sim 2.4$.

\subsection{Uncertainties}\label{sec:uncertainties}

\subsubsection{Ionizing Background Fluctuations}\label{sec:uvbg_fluctuations}

The results presented above all assume a spatially uniform shape and
intensity of the ionizing background, following the convention of all
similar previous studies.  Although we find similar results for two
representations of the {\em mean} background (Haardt \& Madau, and
Faucher-Giguere), we did not explore the uncertainty or scatter that
would be introduced by place-to-place variations in the radiation
field.

This subject has largely been ignored in the previous literature on
IGM abundances, but recent work on the reionization of \hi and \heii
contains arguments that are relevant to observations of heavy element
lines.

Following the development of \citet{fardal_uvbg} and \citet{meiksin},
the variance in the background intensity at a given energy is a
function of the source density of photon emitters at that energy, the
mean free path of photons with that wavelength through the IGM, and
the variance in source spectrum from object to object.  When ionizing
sources are rare, differ in SED from object to object, and photons
have a small mean free path, the radiation seen by a random absorber
tends to be dominated by a single source and will vary significantly
from place to place.

The production of \hi ionizing photons at $z=4.3$ occurs mostly in
galaxies which are relatively numerous, and the mean free path for
these photons is large because \hi is already reionized.  So, many
sources will be contained within an \hi photo-attenuation volume and
the background should be comparatively uniform at 1 Ryd (except in the
immediate vicinity of QSOs).

Overall variations in [C/H] are therefore more sensitive to the \civ
ionization balance, whose atomic transitions are resonant with photons
between $3-30$ Rydbergs.  These photons originate in less numerous
AGN, which could lead to an increased variation in spatial intensity.
Since our high redshift sample is measured at the probable leading
edge of \heii reionization \citep{reimers_2347, 0302_heii, 0302_heap,
  davidsen_kriss_1700, kriss_2347, fechner_1700, shull_2347,
  songaila_siiv}, the IGM opacity from photoelectric absorption must
be considered above the \heii ionization potential (4 Ryd).

For photons of energy $E$, the mean free path is given as
\citep{furlanetto}
\begin{equation}
\lambda = 0.41 \left({1+z}\over{5.3}\right)^{-2} \left({E}\over{4 {\rm
    ~Ryd}}\right)^3 \bar{x}_{\mheii}^{-1} ~{\rm Mpc}
\end{equation}
At $z=4.3$, $x_{\mheii} \approx 1$ since \heii reionization is only in
its earliest phases.  For \civ$\rightarrow$\cv ionizing photons with
4.74 Ryd, this implies $\lambda < 1$ Mpc because of \heii attenuation,
while for \cv$\rightarrow$\cvi ~(24.2 Ryd) the path is $90$ Mpc.  Thus
the most important effect we must consider is stochastic fluctuation
in the 4.74 Ryd background governing the \civ to \cv transition.

At $z=4.3$, simulations indicate that 10-20\% of the universe resides
in ionized ``bubbles'' of \heiii \citep{mcquinn}.  \heii \lya
observations focus only on the interiors of these bubbles, where the
ionized fraction is high and \heii absorption troughs are not totally
saturated.  The lower oscillator strength of \civ renders it
observable in both \heiii ionized bubbles, and the \heii-neutral walls
in between.  In the bubble walls, \heii absorption strongly
attenuates the UV background near 4 Ryd, affecting the ionization
balance of carbon.  Inside of bubbles, stochastic variations in the
proximity of hard photon sources and attenuation by \heii Lyman-limit
systems generates intensity variations both above and below the mean
spectrum.  In the following two sections we consider each of these
environments in separate detail.

\subsubsection{UV background fluctuations in the interior of \heiii bubbles}\label{sec:bubbles}

To study ionization variations of carbon inside \heiii bubbles, we
follow the arguments of \citet{furlanetto}, who develops a simple
Monte Carlo framework suited to the problem.  The method was
originally intended for studying variations in the 4 Ryd background in
\heiii bubbles at $z\sim 3$; a minor set of modifications renders it
useful for studying the \civ$\rightarrow$\civ transition at at
$E=4.74$ Ryd and $z=4.3$.  Essentially all opacities in the model are
scaled down to reflect the smaller cross section of \heii at higher
energy, and fluxes from the ionizing sources is also scaled down to
reflect their SED.

We defer a full description of the method to Furlanetto's paper.  Very
briefly, all quasars are assumed to live inside of \heiiinsp-ionized
bubbles.  The space outside of bubbles is opaque to \heiinsp-edge
photons; inside of bubbles there is still attenuation from the \heii
analog of Lyman limit systems, but between these systems the mean free
path is large.

Because QSOs are rare, the typical bubble contains only one to a few
sources, so a Monte Carlo approach is used.  Ionizing sources are
drawn at random from the luminosity function of \citet{hopkins_lf},
and placed uniformly within each bubble volume (no sources are located
outside of bubbles).  Then, the spatial variation of the background
$J$ is calculated by compiling trial statistics with varying QSO
luminosities, spatial locations in the bubble, Lyman limit
attenuation, and UV spectral indices from the sources.

The Monte Carlo calculation outputs a distribution of
$j=J/\left<{J}\right>$, where $\left<{J}\right>$ represents the mean
value of the background in a fully \heiinsp-ionized IGM.  The $J$
distribution at the \civ$\rightarrow$\cv ionization edge can be used
to examine variations in our \civ ionization corrections.

The spread of $j$ depends strongly on the radii of the \heiii bubbles,
denoted as $R$, since variations diminish once bubbles become large
enough for absorbers to ``see'' multiple QSOs.  
\citet{mcquinn} and \citet{bolton_heii} have shown in simulations that
the distribution of $R$ does not evolve strongly with redshift and is
governed primarily by QSO duty cycle.  Typical values range from
$20-40$ comoving Mpc, with the spread at any redshift larger than the
difference between $z=4$ and $z=2$.  The spread in $j$ also depends on
the photoattenuation length $r_0$, which parameterizes the distance
that \civ ionizing photons travel from their point of origin before
reaching a \heii lyman limit system.
\citet{furlanetto} calculates $r_0$ for \heii ionizing photons at
$z=2,3,4$ and demonstrates that its luminosity-weighted mean evolves
very little.  \civ ionizing photons have slightly higher energies;
their attenuation is still dominated by \heii absorption, but the
relevant $r_0$ increases very slightly because of the smaller \heii
cross section at 4.74 Ryd.

\begin{figure}
\epsscale{1.0} 
\plotone{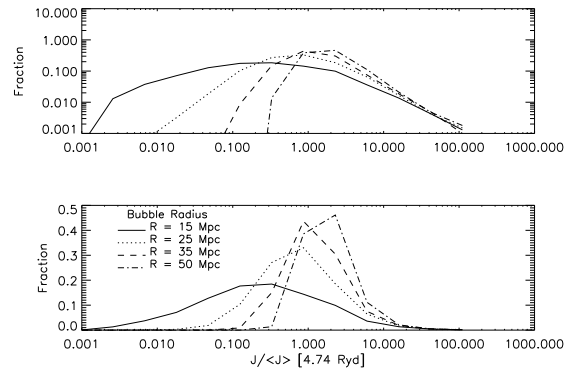}
\caption{Distribution function of flux at the \civ$\rightarrow$\cv
  ionization edge, in \heiiinsp-ionized bubbles at $z\sim 4.3$.  The
  flux is normalized in units of the mean, $\left<{J}\right>$.  Scatter above the
  mean is dominated by proximity to individual QSOs, which enhances
  the background over a wide range of frequencies above 1 Ryd.
  Scatter on the low side arises from stochastic variations in the
  density of nearby QSOs and shielding from \heii patches.}
\label{fig:bubbles}
\end{figure}

Figure \ref{fig:bubbles} shows the distribution of $J$ at $4.74$ Ryd
in logarithmic (top panel) and linear (bottom panel) forms, for
bubbles ranging from $R=15$ to 50 comoving Mpc.  In all cases we have
assumed $r_0=35$ Mpc, as in \citet{furlanetto}.  As expected, the
shape is very similar to what was derived for \heii by
\citet{furlanetto}, except the distributions are slightly narrower
because of the increased transparency of \heii at 4.74 Ryd.  Above the
mean background flux, the curves converge for all bubble sizes.
\citet{furlanetto} observed this same effect for \heiinsp; it
represents environments where the background is dominated by a single
source.  In this case the variance is driven by the shape of the QSO
luminosity function and the random distribution of QSO-absorber
spacings.

According to Figure \ref {fig:bubbles}, variations on the high-side
are limited to a factor of $\sim 10\times$ enhancement in the
background.  Because this enhancement is driven by proximity to a
single QSO, it should lead to an increase in the UV background across
a wide range in frequency.  When examining the \civ ionization balance
in these neighborhoods, we generate an ``enhanced'' toy model of the
background spectrum which boosts the flux for all energies above 1 Ryd
by a constant factor of $5-10$.  Below 1 Ryd the background remains
the same, since it should still be dominated by integrated starlight
from more numerous galaxies.

Below the mean $J$, the distribution in Figure \ref{fig:bubbles} is
more extended and varies with bubble size.  The increased variance for
small bubble sizes results primarily from shot noise in the number of
quasars: by $R=15$ Mpc most bubbles have either no source or one
source, and sources outside the bubble are shielded from view.
This effect is strongest at the \heii ionization edge, and still
significant at 4.74 Ryd.  But the bubble walls are optically thin to
photons with $E<4$ Ryd and also for higher energy photons including
the $24.2$ Ryd \cv$\rightarrow$\cvi edge.  Thus it is appropriate
simply to use the mean UV background spectrum for \hi, as well as all
carbon transitions except \civ$\rightarrow$\cv (and possibly
\ciii$\rightarrow$\civ, see Section \ref{sec:sawtooth}).

To model effects at 4.74 Ryd, we attenuate the \citet{haardt_cuba}
spectrum by an absorbing column with $\tau = \tau_0(E/4 {\rm
  Ryd})^{-3}$.  This produces the desired suppression at $4.74$ Ryd
while preserving the cosmic averaged spectrum in optically-thin
regions.  The normalization $\tau_0$ is adjusted to
produce a factor of 10-100 decrease in the background at 4.74 Ryd,
consistent with the distribution in Figure \ref{fig:bubbles}.

\begin{figure}
\epsscale{1.0} 
\plotone{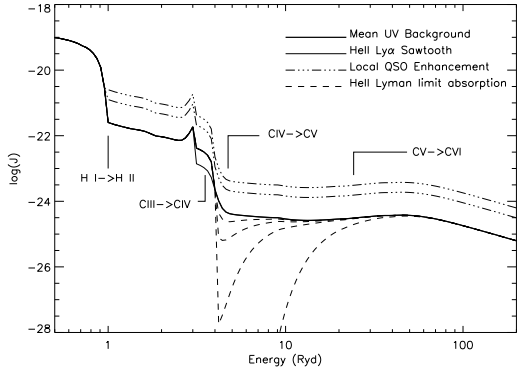}
\caption{Model UV background spectra used to evaluate the effects of
  variations on our inferred [C/H] estimates.  The thick solid line
  represents the mean \citet{haardt_cuba} background.  Curves below
  this mimic the suppression of 4.74 Ryd photons by \heii Lyman limit
  systems (inside of bubbles) or the neutral walls between bubbles.
  Curves above the mean represent the proximity zones of QSOs.}
\label{fig:toymodels}
\end{figure}

\subsubsection{UV background fluctuations in the \heii-neutral inter-bubble medium}

In simulations of \heii reionization, \heiii bubbles only permeate
10-20\% of the IGM by volume, so we must consider the shape of the
spectrum in the \heii walls between bubbles.  This scenario is
discussed in \citet{faucher_uvbg}; essentially flux at the \heii edge
is completely suppressed because in \heiinsp-neutral regions the
optical depth can easily exceed $\tau\sim 100$.  Once again, the
background at lower energies is spatially uniform, being optically
thin after \hi reionization (and, for low energies, dominated by
galaxies).  Likewise at higher energies the \heii cross section
declines and the hard flux recovers to a uniform background.

We therefore model the background spectrum in \heii walls the same as
in the ``suppressed flux'' regions of ionized bubbles: a \citet{haardt_cuba}
standard mean background having \heii absorption blueward of the edge
with $\tau = \tau_0(E/4 {\rm Ryd})^{-3}$.  However in the neutral
walls the normalization of optical depth is much higher.  Whereas
attenuation from Lyman limit systems in bubbles reduces the flux by a
factor of 10-100, the completely neutral bubble walls with $\tau\sim
100$ experience attenuation by 44 orders of magnitude.

\subsubsection{The effect of spatial fluctuations on [C/H] estimates}

\begin{figure}
\epsscale{1.0} 
%\plotone{/Users/Rob/cloudy/z4civ/fluctuations_ionization.ps}
\plotone{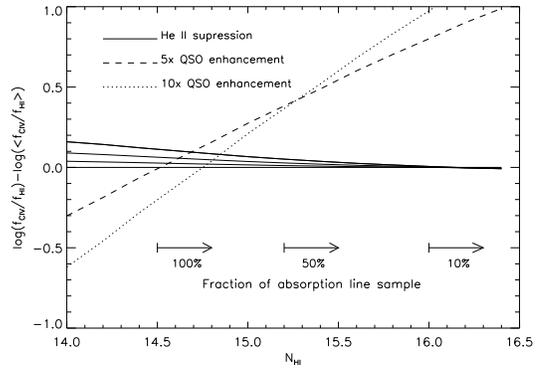}
\caption{CLOUDY calculations showing correction factors to the
  ionization term in Equation 2.  In regions where the 4.74 Ryd
  background is suppressed, little or no change is apparent.  In
  regions of enhanced flux and high density, we may overestimate the
  abundance by up to $\sim 1$ dex.}
\label{fig:chchange}
\end{figure}

Figure \ref{fig:toymodels} shows the range of ionizing background
models we considered when investigating spatial fluctuations.  The
thick solid line represents the mean \citet{haardt_cuba} quasar+galaxy
spectrum at $z=4.3$.  Four models with \heii attenuation of
$\tau_0=1,3,10,100$ fall below the mean.  Smaller values of $\tau_0$
represent conditions on the lower half of the $J$ distribution inside
of \heiii bubbles, the larger values are characteristic of the \heii
bubble walls.  Above the mean curve are two models with enhanced hard
UV flux (i.e. proximate to QSOs), shown in dot-dashed lines.

We ran each of these spectra through our CLOUDY model grid to examine
the change to the ionization fraction as a function of column density.
Figure \ref{fig:chchange} shows the result of this exercise, where we
plot the change in ionization correction for each different choice of
background spectrum.  A value of zero indicates no change in [C/H] for
a given line; a positive value implies a {\em lower} metallicity for a
given set of \civ and \hi measurements.

The solid lines near zero represent the four ``attenuated'' models
with varying optical depth at 4 Ryd, increasing upwards.  For these
models the \hi ionization fraction does not change at all because the
universe is already optically thin to 912 \AA ~photons at $z\sim 4.3$
and the background is dominated by flux from numerous galaxies.  The
\civ ionization fraction does not change either, which results in a
very small total correction of $\lesssim 0.2$ dex to [C/H] across our range
of $\nhi$.  

The small change in $f_{\mciv}$ is surprising at first glance given
that the flux at the \civ$\rightarrow$\cv ionization edge changes by
many orders of magnitude.  The reason this effect is so small is that
even for the mean spectrum, over 50\% of the carbon is already in the
\civ state.  So, a reduction in the \civ$\rightarrow$\cv rate---which
would increase $f_\mciv$---cannot change the \civ fraction by more
than a factor of $\sim 2$.

A much larger correction is obtained when considering upward
fluctuations in the UV background from proximity to bright quasars.
While this change can be significant for selected systems, the number
of affected data points should be small.  Major upward fluctuations
occur only in the largest bubbles surrounding rare, luminous QSOs, and
even in these bubbles the probability of lying in a proximate region
is low according to Figure \ref{fig:bubbles}.  Such an error would
push our abundances {\em lower}, enhancing the evolutionary trend
discussed in Sections \ref{sec:qualitative} and \ref{sec:km_metals}.

\subsubsection{\heii Sawtooth}\label{sec:sawtooth}

\begin{figure}
\epsscale{1.0}
\plotone{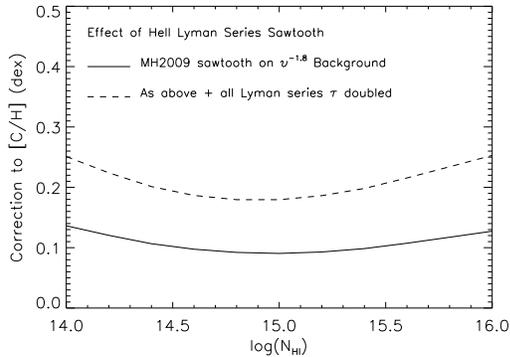}
\caption{Effect of flux suppression at the \ciii ionization edge from
  the \heii \lya forest ``sawtooth'' spectral modulation.}
\label{fig:sawtooth}
\end{figure}

\citet{HM_sawtooth} have recently emphasized the importance of a
``sawtooth'' imprint left on the ionizing background spectrum between
3-4 Ryd, from line absorption of the \heii ``Lyman'' series.  This
relates to the \civ balance because the transition from \ciii to \civ
occurs at 3.5 Ryd.  These authors' updated model of the UV background
including the sawtooth modulation leads to a decrease in flux at the
\ciii ionization edge of 0.2-0.3 dex at $z=3$.  They also explored
models where an artificially delayed \heii reionization led to a
reduction in flux at the \ciii edge of over a full dex.  This latter
case may be more appropriate for studying the $z=4.3$ IGM since \heii
reionization should not be complete at this epoch and the resulting
opacity may be quite high.  Statistical studies of metal lines at high
redshift provide some evidence for this effect \citep{agafonova}

For the mean spectrum with no \heii sawtooth, the majority of carbon
is in the \civ state at $z\sim 4.3$.  Therefore, in principle a
softening of the background should shift the balance toward \ciii and
\ciinsp, reducing $f_\mciv$ and increasing [C/H] estimates in the
process (Equation 2), thereby reducing the evolution signal.

To explore this effect, we obtained a copy of the sawtooth spectrum
from F. Haardt, and used it to recalculate ionization corrections.  In
its raw form, the sawtooth model has a very hard spectrum that is
disfavored by IGM abundance studies at lower redshift
\citep{schaye_civ_pixels, simcoe2004}.  It produces an abundance
gradient that decreases with density, leaving very high abundances in
weak \lya forest lines.

Rather than use this form as-is, we instead examined the ratio of the
sawtooth spectrum (HM+S in their Figure 1) to a separate model
calculated with identical input parameters but no sawtooth included
(HM in their Figure 1).  This isolates the effect of the \heii Lyman
series, which we then multiply into the softer background spectra
shown in Figure \ref{fig:uvbg}.  Clearly this is not a self-consistent
way to model \heii absorption in the IGM, and \citet{HM_sawtooth}
point out that even their models do not fully capture the patchy
nature of \heii reionization that could be important here.  But
lacking a full simulation suite, our approach captures the spectral
shape of the sawtooth, and provides an heuristic tool for examining its
impact on our results.

Figure \ref{fig:sawtooth} illustrates the change in [C/H] resulting
from use of our modified sawtooth spectrum, as a function of $\nhi$.
The solid line shows the MH09 model; for the dashed curve we
arbitrarily increase the optical depth across the sawtooth by a factor
of 2.  Depending on the degree by which the spectrum is changed, the
correction can be anywhere from 0.08 to 0.2 dex, and always in the
upward sense.  The sawtooth modulation therefore will decrease the
evolution signal between our $z=4.3$ and $z=2.4$ samples.

\begin{figure}
\epsscale{1.0} 
%\plotone{/Users/Rob/z4civ/civ_fits/km_saw_20110327.ps}
\plotone{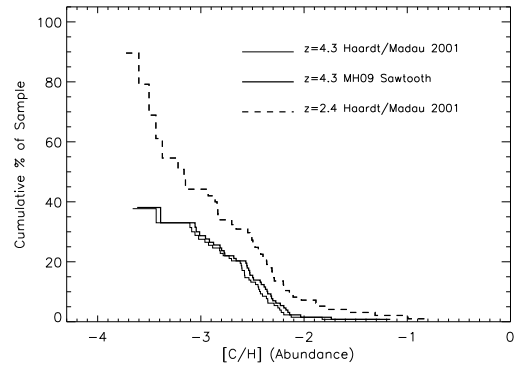}
\caption{Comparison of Kaplan-Meier distributions for the standard
  $z=4.3$ \citet{haardt_cuba} model, the $z=2.4$ standard model, and
  the $z=4.3$ model incorporating a 1 dex decrement at 3-4 Ryd from
  the sawtooth modulation.  This modification changes the shape of the
  Kaplan-Meier distribution, such that the high [C/H] regions do not
  evolve, but the low [C/H] regions do.}
\label{fig:km_saw}
\end{figure}

Figure \ref{fig:km_saw} shows the Kaplan-Meier distribution in [C/H]
obtained by running our survival analysis with the HM+S background at
$z=4.3$.  The two comparison curves show the distributions at $z=2.4$
and 4.3 for the unmodulated spectrum.  For our modified HM+S spectrum,
the [C/H] distribution shifts uniformly to the right by $\sim 0.06$
dex; a similar plot using the background with $2\times$ inflated \heii
optical depths would shift $\sim 0.15$ dex higher than the
distribution from the unmodulated spectrum.

Recall that we observe a 0.5 dex change in [C/H] between $z=4.3$ and
$z=2.4$; if one forces [O/C]=0.5, or requires a strong correlation
between density and metallicity, the evolution could be as small as
0.3 dex.  A very strong sawtooth modulation in the UV background could
in principle remove 0.15 dex of our evolution signal.  If one applies
both the sawtooth spectrum {\em and} forces abundances to [O/C]=0.5,
it is possible to derive a solution with only 0.10-0.15 dex of
evolution in the carbon abundance, much closer to the no-evolution
scenario.  This situation arises only for a set of assumptions
stressed to a particular end of our parameter space, but its
possiblility should not be ignored.

\begin{figure}
\plotone{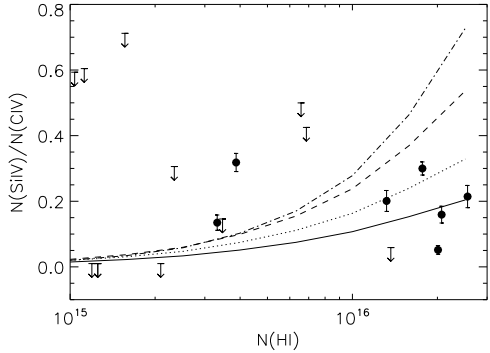}
\caption{\siiv/\civ ratio for systems in the sample with detected
  \civ.  Model curves show the predicted trends for various degrees of
  sawtooth absorption in the UV background spectrum.  The solid curve
  incorporates no sawtooth, dotted line is the \citet{HM_sawtooth}
  form modified as described in the text.  The dashed and dot-dash
  lines are for sawtooth spectra with optical depth arbitrarily
  multiplied by $2\times$ and $4\times$, respectively.  These models,
  which would be needed to substantially shift the [C/H] distribution,
  are disfavored by the few measurements shown here.}
\label{fig:sivciv}
\end{figure}

The chief effect of the 3-4 Ryd decrement at $z\sim 4.3$ is to take
carbon from the \civ state and move it to \ciiinsp, so we should in
principle be able to see an increase in the \ciiinsp/\civ ratio using
the \ciii 977\AA ~line, or even an increase in \ciinsp/\civ at
1334\AA.  We searched our data for systems where this measurement
could be made, but unfortunately the results do not provide strong
constraints on the magnitude of the sawtooth effect.  The \ciii lines
tend to be blended with \lya forest absorption, and \cii lines are too
weak to detect in all but our strongest few systems.  One of the
strongest constraints may come from pixel-optical-depth analyses.  The
\ciiinsp/\civ ratio is not a good diagnostic at $z\gtrsim4$.  However,
the models in \citet{schaye_civ_pixels}---which show no trend of [C/H]
with redshift---may yield a declining cosmic abundance from high
values in the past toward low values at present using a sawtooth
background---a solution we are generally prejudiced against.

The \siiv/\civ ratio is another diagnostic of spectral hardness which
is accessible for a limited number of systems in our sample.  Figure
\ref{fig:sivciv} shows the calculated values, or upper limits (where
no \siiv is detected).  The solid lines show predicted values of
$N_\msiiv/N_\mciv$ for different prescriptions of the sawtooth,
calculated using CLOUDY with Solar relative abundances
\citep{grevesse_solar_abund}.  At high $\nhi$ where the most reliable
measurements are located, the data are broadly consistent with the
non-sawtooth spectrum and our modified \citet{HM_sawtooth} sawtooth
form.  Artificially boosted sawtooth absorption overpredicts the amount
of \siiv seen in these systems, although the number of points is too
small to draw broad conclusions.  This diagnostic suffers a slight
degree of ambiguity because of possible non-solar values of [Si/C].
However many absorbers at high redshift show [Si/C]$>0$, in which case
there would be an even stronger conflict for the heavily absorbed
sawtooth spectra.

\subsubsection{Continuum Fitting}\label{sec:contiuum}

At high redshifts an additional possible bias is introduced from
uncertainties in the absolute continuum flux from the background
quasar over the \lya forest region.  We have followed the typical
procedure of estimating continuum levels by fitting a low-order spline
across relatively unobscured regions of the forest.  This procedure
becomes increasingly inaccurate at high redshifts, as the opacity of
the \lya forest increases, and it becomes increasingly difficult to
find regions with transmission of unity.

\citet{faucher_continuum} present a treatment of this same problem in the
context of \lya forest opacity measurements.  Drawing mock spectra
from \lya forest simulations at various redshifts, they performed
manual continuum fits similar to the ones described here and examined
the resultant deviation from the known, true continuum as a function
of redshift.  Their results indicate that at $z=4.2$ (our sample
redshift) a manual estimate could systematically place the continuum
too low, by $\sim 17\%$.  By comparison, at $z=2.5$, where \ovi and
other studies focused, the correction is of order $2\%$.

To examine this effect, we created a ``continuum corrected'' version
of our highest quality sample spectrum (BR0353-3820) and refit \hi
column densities to the full \lya forest.  Following the empirical
approach of \citet{faucher_continuum}, we adjusted the
continuum-normalized flux at each wavelength in the \lya forest
downward by a factor of $(1-C(z))$, where $C(z)=1.58\times
10^{-5}(1+z)^{5.63}$, to account approximately for the bias introduced
by incorrect continuum fits.  We then re-calculated $\nhi$ at each
redshift for our full sample.

\begin{figure}
\epsscale{1.0}
\plotone{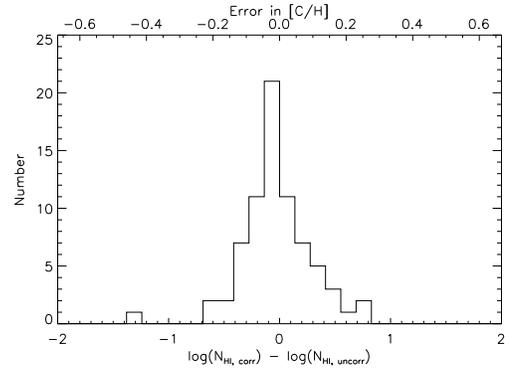}
\caption{Error introduced in $\nhi$, and hence [C/H], caused by errors
  in our estimate of the QSO continuum in the \lya forest.  No
  systematic trend is apparent, and most of the scatter is at the
  $\sim \pm 1$ dex level.}
\label{fig:contin}
\end{figure}

Figure \ref{fig:contin} shows the results of the re-fitting exercise
for BR0353-3820, where as before we use multiple Lyman series
transitions to constrain $\nhi$.
Somewhat surprisingly, we found that the column density for the median
system was nearly unchanged, or at least was offset by an amount
smaller than the $\sim \pm 0.2$ dex point-to-point scatter.  This may
result from the fact that in each system, the $\nhi$ measurement is
driven by that Lyman series line which is closest to saturation but
not bottomed-out across the profile.  The fractional error introduced
by poor continuum fits is smallest at low flux levels so by choosing
the right transitions this is minimized.  At least it seems clear that
changes in the continuum level at $z\sim 4$ lead to errors at the
level of $\lesssim 0.5$ dex in $\nhi$.

In general, continuum errors enter into [C/H] in two ways.  One is a
simple error in $N_\mciv/N_\mhi$.  However, because $f_\mciv\propto
N_\mhi^{2/3}$ (Figure \ref{fig:fhi}), we also underestimate the
neutral fraction, which partly counteracts the misestimate of
$N_\mciv/N_\mhi$.  Thus the error in [C/H] is only $\frac{1}{3}$ the
logarithmic error in $\nhi$---in our case, $\sim 0.15$ dex of random
scatter with no systematic offset.

There is an additional, much smaller contribution from an incorrect
estimate of $f_{\mciv}$, which is derived from $\nhi$.  Over the
column density range of our sample, the \civ fraction is near its
peak.  This results in a broad minimum in the logarithmic derivative
of the \civ fraction with $\nhi$.  For the \citet{faucher_uvbg}
background (as a representative example; see Figure
\ref{fig:fciv_z4}),
\begin{equation}
\left|{{{d \log(f_{\mciv})}\over{d\log (\nhi)}}}\right|\lesssim 0.1
\end{equation}
from $14.5<\log (\nhi) < 16.5$, with the derivative near zero between
15.0 and 15.75.  The \civ column density in Equation 2 is not
sensitive to continuum fitting errors and remains constant.

Taken all together, the corrections to $\nhi, f_{\mhi},$ and
$f_{\mciv}$ would lead to a combined correction in [C/H] of only
0.032-0.034 dex for each 0.1 dex change in $\nhi$.  Moreover the sense
of the change is such that the measured abundance at $z=4.3$ becomes
{\em lower} as $\nhi$ is increased, which tends to enhance the trend
of declining carbon abundance with increasing redshift.

\subsection{Summary of systematic uncertainties}

To synthesize the results of this section: the main systematic sources
of uncertainty in our abundance estimates come from the UV background
used to calculate ionization corrections, and the determination of \hi
column densities in the thick \lya forest at $z\sim 4.3$.  We have
tried two choices for the ionizing background: the commonly-used
\citet{haardt_cuba} model with quasars + galaxies, and a newer
calculation of \citet{faucher_uvbg}.  Both versions of the spectrum
yield an evolving carbon abundance with redshift, with the
\citet{faucher_uvbg} version giving a slightly larger derivative.  

We investigated the effect of spatial variations in the UV background
due to photoelectric absorption from \heii, and stochastic fluctuations
in the ionizing source population.  These variations do not affect the
\hi neutral fraction, or transitions relevant to \civ other than the
\civ to \cv transition at 4.74 Ryd.  Near this edge, the background
can vary by factors of 10-100 inside of \heiiinsp-ionized bubbles, and
many orders of magnitude in the neutral bubble walls.  However, CLOUDY
tests revealed that changes in the 4.74 Ryd flux, absent changes at
other wavelengths, affects our abundances only at the 0.1 dex level.
It is possible that we have overestimated abundances for systems in
the proximity zones of bright quasars by $\sim 1$ dex or more, but
these systems are quite rare and should not drastically alter our
distributions.  

The inclusion of ``sawtooth'' absorption from the \heii Lyman series
at 3-4 Ryd leads to a systematic increase of $0.06$ dex in our median
[C/H], for an absorption strength consistent with the results of
\citet{HM_sawtooth}.  If the sawtooth absorption strength is
artificially inflated this offset becomes correspondingly larger.

Abundance errors from a systematic underestimate of $\nhi$ (as would
be found from continuum fitting errors) are limited to $\lesssim 0.1$
dex, because for every 0.1 dex increase in $\nhi$, there is a
corresponding increase of $0.07$ dex in $f_\mhi$ offsetting the
effect, and little change in the \civ ionization fraction.

Taken together these tests suggest that [C/H] abundances at $z\sim
4.3$ are less sensitive to systematic errors than one might naively
expect.  Different choices of the mean background will change the
overall median abundance but do not affect our conclusion that the
carbon abundance is evolving except under stressed sets of
assumptions.  Most of the effects studied above tend to {\em enhance}
the evolutionary signal, if we have underestimated their importance.

\section{Discussion}\label{sec:discussion}

\subsection{The Integrated Carbon Production Between $z=4.25$ and $z=2.4$}\label{sec:c_production}

The the total mass flux of carbon into the \lya forest may be
estimated simply from our data by integrating the abundance-weighed
\hi column density distribution.  As described in \citet{simcoe2004},
the contribution of carbon atoms (in all ionization states) to the
closure density may be estimated as:
\begin{eqnarray}
\Omega_{C}={{1}\over{\rho_c}}\left({{C}\over{H}}\right)_\sun\left<{10^{[C/H]}}\right>\left({{c}\over{H_0}}\right)^{-1}\times \nonumber\\
\int{N_{\mhi}f_{(\nhi,X)}dN_\mhi}
 \end{eqnarray}
where $\rho_c$ is the (current) closure density, and $(C/H)_\sun$ is
the solar carbon abundance by number. The \hi column density
distribution $f_{(\nhi,X)}$ is defined as $d^2{\cal N}/dXdN_{\mhi}$
with ${\cal N}$ being the number of absorbers in a survey of comoving
pathlengh $dX$ (Shown in Figure \ref{fig:cddf}).

The mean carbon abundance, shown in angled brackets, may be derived
from the distributions in Figure \ref{fig:km_ch}.  Assuming a lognormal
parameterization of the abundance, the mean may be determined as
\begin{equation}
\left<{10^{[C/H]}}\right> = \exp \left[{\ln10\left<{\left[{{C}\over{H}}\right]}\right>+\frac{1}{2}(\ln 10\times \sigma)^2}\right]
\end{equation}

\noindent For our measured median abundance of $[C/H]=-3.55$ and scatter
of $\sigma=0.7$ dex, this amounts to a mean carbon abundance of
$10^{-2.81}$ at $z\sim 4.3$.  The same equation applied to $z\sim 2.4$
yields a mean abundance of $10^{-2.36}$ at lower redshift.

We substitute these values into Equation 8 to derive
$\Omega_{C}=2.7\times 10^{-8}$ at $z=4.3$, and $4.8\times 10^{-8}$ at
$z\sim 2.4$.  This represents a factor of 1.7 increase over an
interval of just 1.3 Gyr.  Put another way, {\em roughly half of the
  carbon observed in the \lya forest at $z\sim 2.4$ was deposited into
  the IGM after $z\sim 4.3$.}  Even in the conservative limit where
0.2 dex of the difference between our redshift points is attributed to
density effects, the fraction of ``recently deposited'' carbon would
be about one-third of the total at $z\sim 2.4$.  The rise in
intergalactic abundance suggests a connection with the concurrent rise
in the cosmic star formation rate and/or the luminosity density of AGN
in this same redshift interval.  Many of these metals may reside in
enriched regions near high redshift galaxies, given their relatively
recent production.

Outflows are of course commonly seen in the spectra of star
forming galaxies at these redshifts \citep{lbg_winds, steidel2010,vvds_lbg}
and there is evidence of spatial association between strong \civ
systems and UV-selected galaxies at $z\sim 3$ \citep{kurt_winds,
  steidel2010}.  Our measurement would seem to connect these
correlations in the local galactic environment to the global growth
of metal abundance in the IGM, in a temporal sense.

Estimates of the stellar population ages for $z\sim 3$ galaxies
typically range between $100-200$ Myr to $\sim 1 $ Gyr
\citep{alice_spitzer}.  This means that the archetypal Lyman Break
galaxy at $z=2.4$ could be built from scratch within the redshift
interval of our \civ survey.  The winds seen in LBG spectra travel at
several hundred km/s, so even if these galaxies began driving totally
unbound winds at birth, the wind material could only have traversed
several hundred kpc and not deep into the IGM.

The \civ column densities studied here ($\nciv \sim 10^{12}$)
generally fall below the range where correlations with galaxies are
strongest \citep[$\nciv>10^{13.5}$---][]{kurt_winds, steidel2010}.
Moreover \citet{songaila_winds} studied similar \civ systems to the
ones presented here, concluding that the weaker lines dominating
absorption selected samples are too quiescent to trace actively
evolving wind bubbles.
However they do trace densities that would be found in the filaments
seen in cosmic simulations, where galaxies also reside.  These
filaments may be populated with moderately recent, cooled wind relics
that would manifest as the weaker \civ systems seen in our survey and
other sensitive \civ searches in the \lya forest, although further
theoretical work would be required to assess if this scenario holds up
in detail.

\subsection{Comparisons with the Global Star Formation Rate}\label{sec:sfr}

For any moment in time, the difference between the intergalactic
carbon growth rate and the yield-weighted star formation rate measures
the global efficiency of feedback processes at ejecting metals from
nascent galaxies.  Observations of individual star forming galaxies in
the local universe indicate that galactic winds can be significantly
loaded in mass and especially in metals \citep{crystal_wind_yield}.
During periods of intense star formation, it is possible that all
metals produced in a burst will escape; in more quiescent periods
these same elements would settle back into the galaxy's ISM.  At
$z=2.5-4.5$ individual galaxy measurements are difficult, but using
our \civ sample we may estimate a globally averaged efficiency of
metal ejection.

First, we calculate the total stellar mass formed over our redshift
interval by integrating the cosmic star formation rate density of
\citet{reddy}.  
Over the interval $2.4<z<4.3$ ($\Delta t=1.34$ Gyr), this yields a
total formed stellar mass of $1.6\times 10^{8} M_\sun$/Mpc$^3$, or an
average of about 0.16 $M_\sun$ Mpc$^{-3}$yr$^{-1}$.  We then estimate
the carbon output resulting from this star formation activity using
yields from \citet{woosley_weaver_yields}\footnote{The IMF correction
  from UV luminosity to stellar mass is counterbalanced by the
  IMF-weighting of the yield, and it is the UV-bright stars which
  matter most for enrichment.  As long as the same IMF is applied
  consistently for both calculations the detailed form matters less
  for metal production.  We thank the referee for raising this
  point.}.  The IMF-weighted carbon yield, calculated assuming a
Salpeter IMF from 0.5 to 40 $M_\sun$ and for solar metallicity, is
$1.2\times 10^{-4}$.  The yield for progenitors with $M<10M_\sun$
(which mostly still lie on the main sequence) is set to zero.  Folding
this into the star formation rate, we find that the total production
of carbon from core collapse supernovae between $z=4.3$ and $z=2.4$ is
\begin{equation}
\Delta M_C \approx 2.1\times 10^4 M_\sun ~{\rm Mpc}^{-3} ~{\rm (Produced ~in
  ~galaxies)}.
\end{equation}

In the previous section, we found that the total growth of
intergalactic carbon came to $\Delta \Omega_{\rm C}=2.2\times10^{-8}$,
or
\begin{equation}
{{d}\over{dt}}\Omega_{\rm C} \sim 1.9\times10^{-17} {\rm yr}^{-1}.
\end{equation}
Multiplying by the critical density, we obtain the integrated
volumetric increase in carbon mass: 
\begin{equation}
\Delta M_{\rm C}=7.1\times 10^3 M_\sun ~{\rm Mpc}^{-3} ~{\rm(Added ~to ~IGM)}.
\end{equation}
A simple comparison of Equations 8 and 10 shows that the rate of
carbon enrichment in the IGM is roughly a factor of $\sim 3$ lower
than the rate of carbon production in stars at this epoch.  This
analysis is subject to substantial uncertainties in calculating the
SFRD, IGM carbon abundance, and yields.  It is somewhat remarkable
that in spite of this, the carbon production and pollution rates are
of similar magnitude.  Taken at face value, the result suggests that
galaxies at $z\sim 2-4$ keep roughly $70\%$ of the heavy elements they
produce, and donate the other $30\%$ back to the IGM.

There is precedent for this behavior at both low and high redshift.
At the high redshift end, \citet{simcoe2004} estimated galaxy yields
based on a closed-box chemical enrichment model of the \lya forest.
This analysis relates to the present one, except that it is an
integral calculation whereas the present version is differential.
However the results are similar: the closed-box model requires that
galaxies lose at least $\sim 10\%$ of their heavy element mass to
enrich the IGM to observed levels by $z\sim 2.4$, and possibly more.

Our escape fraction of $\sim 30\%$ is also comparable to those derived from
very limited samples of dwarf starbursts with outflows in the
local universe.  \citet{crystal_wind_yield} report ejection
efficiencies of $15-20\%$ for the dwarf starburst NGC1569.  For the
local prototype starburst M82, \citet{strickland_m82} report a star
formation rate of 4-6$M_\sun$ yr$^{-1}$ with mass outflow rates of 1-2
$M_\sun$ yr$^{-1}$.  If the SFR is weighted by yield and the ejecta
have abundances above $\frac{1}{10}Z_\sun$ (observations suggest
$Z\sim 5Z_\sun$ in the wind), the metal outflow rate even exceeds the
metal production rate, presumably because of mass loading from the
ISM.

While selected individual galaxies in the local universe have mass and
metal outflow rates similar to these values, the high redshift growth
in abundance would require that essentially every galaxy drive
outflows with this efficiency.  

\subsection{Metal Mixing and the Smoothing Scale for Abundance Measurements}

The abundances measured in this work---as well as in
\citet{simcoe2004} and \citet{schaye_civ_pixels}---are understood to
be smoothed on scales comparable to the Jeans length for the
structures in question, which fall in the range $\sim 50-150$ kpc.
\lya forest clouds have small overdensities relative to the cosmic
mean, and are generally thought to arise in structures that are
marginally Jeans unstable \citep{schaye_forest}.  This sets the
relation between $\nhi$ and $n_H$ seen in simulations, which we have
exploited in calculating our ionization corrections.  It indicates
that the sizes, densities, and \hi ionization fractions employed here
should accurately represent the true physical conditions inside of \hi
absorbers.

However the mixing of \civ on $\sim 100$ kpc scales is neither
well-mapped observationally or well-resolved in numerical simulations.
The information that we do know suggests that \civ is not smoothly
distributed throughout \lya forest clouds.  This is suggested
observationally by the slight velocity offsets often observed between
\hi and \civ lines at similar redshift.  Moreover observations of
starburst outflows in the local universe show a highly non-uniform
geometry.  Typically there is a uniform hot X-ray halo powered by the
feedback source, but cooler gas that would be seen in absorption is
often clumpy or filamentary.

Several recent hydrodynamic codes have explicitly tracked the fate of
metals flowing out of galaxies, and these also tend to find a very
non-uniform distribution, with metal absorption coming from small,
higher density regions embedded in the more diffuse \hi medium
\citep{cen_enrichment_sims,oppenheimer}.  However these simulations
employ SPH or other particle based methods, tracking heavy element
mixing by tagging outflow particles.  These methods will underestimate
the diffusion of metals on scales at or below that of a particle,
which are crucial to the mixing process.

Several authors have studied the sizes of \civ absorbing regions along
the line of sight using ionization modeling \citep{simcoe2004}, or
using transverse separation provided by gravitational lenses
\citep{rauch_civ_lens}.  Invariably these studies find small
absorber sizes of $\sim 1$ kpc or less, sometimes much smaller.  The
extreme argument for non-uniform mixing is presented by
\citep{schaye_mixing}, based on fitting heavy element components on
the wings of larger \civ profiles.

\begin{figure}
\plotone{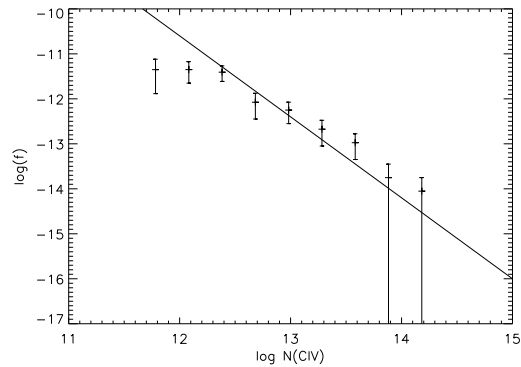}
\caption{Comparison of the \civ column denisty distribution from
  sample spectra with the redshift-independent power law determined by
  \citep{songaila_omegaz}.  For this plot only, we use Songaila's
  cosmology for direct comparison.  Small number statistics set in at
  $N_\mciv \gtrsim 10^{14}$, while incompleteness is apparent at
  $N_\mciv\lesssim 10^{12.5}$.  The few strong systems missed have
  minimal impact on the survival analysis; the statistics are
  dominated by the more numerous, low column density absorbers.}
\label{fig:civ_cddf}
\end{figure}

Using simple scalings, \citet{schaye_mixing} examine the fate of
non-uniform, high metallicity clumps outside of galaxies.  If these
systems start out as hot clumps from superwinds, they are
overpressurized with respect to the intergalactic surroundings, and
will expand until they reach pressure and temperature equilibrium with
the ambient medium.  Once in equilibrium, the metal-rich patch also
reaches a common density with the IGM.  This process occurs on
timescales of 1-10 Myr, essentially instantaneously with respect to
the Hubble time or even the star formation age of galaxies at $z\sim
2-4$.

Small patches of high metallicity embedded in the \lya forest clouds
may therefore dominate the \civ absorption profile, but the \hi
profile is dominated by the extended, metal poor \lya forest absorber.
Because their densities are the same once equilibrium is reached, the
ionization correction for \civ (which depends on $n_H$ derived from
$\nhi$) should actually still be accurate for calculating $f_\mciv$.
This implies that we have correctly measured the relative amounts of
hydrogen and carbon for calculating [C/H].  However this estimate is
essentially averaged over the whole \lya absorber.  The actual
abundance in any single volume element may be either substantially
lower or higher depending on the degree of mixing.

\subsection{Comparison with Previous Studies}\label{sec:thepast}

A number of authors have studied the evolution of \civ at $z>2$.
\citet{songaila_new_civ} notably evaluated $\Omega_{\mciv}$ and the
\civ column density distribution function from $z=2.1$ to $z\sim 6$.
A key finding of this work was the lack of evolution observed in the
\civ CDDF.  Our results are broadly consistent with this result.
Figure \ref{fig:civ_km} illustrates that our data yield a very similar
column density distribution of \civ at redshifts $2.4$ and $4.3$.  The
Kaplan-Meier distribution is somewhat different than a standard CDDF
in that we have not scaled by redshift pathlength, but by reporting as
a percentage of the sample the different pathlengths of the samples
normalize out.  Also, the KM distribution better captures information
about non-detections, which explains why \civ is detected in only
about $50\%$ of our sample (which in turn consists of stronger \lya
forest lines).

To test whether the \civ systems found in our sightlines are
consistent with other surveys from the literature, we show the classic
\civ CDDF in Figure \ref{fig:civ_cddf}, made from a count of
(\civ-selected) absorbers in our spectra.  The power-law fit from
\citet{songaila_omegaz} is shown as a solid line, this fit agrees with
our data over the range $12.4<\log (N_\mciv)<14.0$.  Both we and
Songaila are substantially complete in the range
$13<\log(N_\mciv)<14$.  A straight sum of systems in this $N_\mciv$
interval yields $\Omega_\mciv=1.3\times 10^{-8}$ for our sample at
$z=4.3$, while her result (scaled to $\Omega_M=0.3$) yields $1.1\times
10^{-8}$, in good agreement.

Using her full sample with $13.0<\log(N_\mciv)<15.0$,
\citet{songaila_new_civ} estimates that $\Omega_{\mciv}=1.8-2.7\times
10^{-8}$ over our redshift range.  In Section \ref{sec:c_production},
we calculated the total {\em ionization corrected} $\Omega_{C}$ for an
\hi-selected sample, with result $2.2\times 10^{-8}$ and $4.8\times
10^{-8}$ at $z=4.3, 2.4$, respectively.  From Figure
\ref{fig:fciv_z4}, we see that the \civ fraction in our sample ranges
from $35-50\%$ at $z=4.3$ to $17-36\%$ at $z=2.4$.  These imply an
approximate $\Omega_{\mciv}\sim 1.0-1.5\times 10^{-8}$, on the low
side of Songaila's range but generally consistent.  It is not
surprising that our estimate made in this alternate way is slightly
low, since about half the signal in raw measurements of $\Omega_\mciv$
is produced by the rare, strong systems (i.e. outliers) that we have
not included in our statistical calculation.  The strong systems
picked up in $\Omega_\mciv$ are not well sampled here and may reside
in circum-galactic environments that are locally enriched.

\citet{schaye_civ_pixels} performed a systematic investigation of \civ at
similar redshifts and, importantly, included ionization corrections in
their analysis.  Unlike our method, which relies on measurements of
line column densities and upper limits, \citet{schaye_civ_pixels} performed
a pixel-optical-depth (POD) analysis and compared with forward-modeled
simulations of the IGM to infer [C/H] versus redshift.

The POD method yields a median [C/H]$=-3.47$ at a pathlength-weighted
mean redshift of $3.1$.  This lies between our estimate of [C/H]=-3.1
at $z\sim 2.4$ and [C/H]=-3.6 at $z\sim 4.3$.  A simple linear
interpolation between our two redshift points would yield [C/H]=-3.3
at $z=3.1$, about 50\% higher than the POD estimate.
Schaye et al. find no statistically significant evidence of redshift
evolution, in contrast to our result.

Some of this discrepancy may reflect a real difference in the
measurements and ionization corrections.  But there are several
factors to consider when comparing the two.  First, while we and
Schaye have both employed the updated Haardt \& Madau model of the UV
background with galaxies included, we have softened our spectrum above
10 Ryd at $z=4.3$, to avoid a rise in the X-ray background that would
be inconsistent with the observed decline in the space density of AGN.
Schaye et al noted that in experimenting with models softened above 4
Ryd (simulating \heii reionization), a transition in the middle of
their redshift range would result in a positive detection of
evolution.  For reference, when we use the updated calculations of
\citet{faucher_uvbg} which are {\em not} artificially tuned but which
incorporate \heii ionization into the mean calculation, we measure an
even {\em stronger} evolution by about 0.3 dex.

A second point is that the present study (by design) has its
pathlength weighted at the two endpoints of the redshift interval
where we have detected a time derivative in the metallicity.  The
Schaye et al sample has its pathlength weighted toward the middle of
its redshift range (around $z\sim 3.0$) to maximize signal-to-noise,
and therefore has somewhat less leverage for measuring evolutionary
trends.  For example, of the 19 QSOs in their sample, only 2 have
pathlength above $z=3.6$, and only one has pathlength above $z=4.0$.
Indeed, the relative lack of high redshift, high resolution data in
the Schaye sample was a motivating factor for obtaining our MIKE data
set.

This effect manifests partly as a reduced signal-to-noise for
measuring evolution.  However, it is also the case that the lowest
densities probed in any large sample spanning $\Delta z >1$ will be
found at higher redshift, because of the reduced ionization of the
IGM.  Likewise the highest redshifts are best for measuring low [C/H]
because of the large $f_\mciv$ at earlier times.  Since density,
abundance, and redshift can be artificially correlated, a large
regression analysis like the POD technique may encounter degeneracies
along these basis vectors.  The POD analysis projects a strong signal
along the density axis with weak redshift evolution.  By studying
slices of samples with individual absorbers, we favor a slightly
weaker correlation with density, and a stronger evolution with
redshift.

Finally, we note that simulations of metal enrichment in the IGM have
improved substantially over the last several years.
\citet{oppenheimer} and \citet{cen_enrichment_sims} have performed
extensive studies of metal enrichment in cosmological simulations with
particular attention to carbon and the \civ ion.  These represent
improved updates to the early work of \citet{aguirre_outflows}.  They
find that the total budget of heavy elements in the IGM increases by a
factor of 2.7, or 0.4 dex over this range.  This is very closely in
line with the change that we observe, indicating that wind models
motivated by local observations--but implemented at high
redshift--provide a good match to the trends seen in the carbon
abundance.

\section{Conclusions}\label{sec:conclusions}

We have derived the distribution function of carbon abundance in \lya
forest clouds at $z\sim 4.3$, using a set high signal-to-noise ratio
spectra taken with the MIKE echelle spectrograph on the Magellan Clay
telescope.  The \civ column density or its upper limit was measured
for an \hi-selected sample of 131 discrete absorbers with $\nhi >
10^{14.5}$ cm$^{-2}$, corresponding to $\oden \ge 1.6$.  These
measurements were converted into [C/H] abundances via
density-dependent ionization corrections.  The [C/H] distribution was
then determined via the Kaplan-Meier product limit estimator for
censored data.  Our main findings are that:
\begin{enumerate}
\item{Over the range we can probe, the \civ distribution at $z\sim
  4.3$ appears to be crudely lognormal, with a median of [C/H]=-3.55
  and scatter of $0.8$ dex.}
\item{The median abundance at $z\sim 4.3$ is about 0.3-0.5 dex lower
  than at $z\sim 2.4$, at fixed cosmic overdensity $\rho/\bar{\rho)}$.
  The range quoted reflects differing assumptions about the UV
  ionizing background, and the gradient of abundance with density.}
\item{We examined several sources of uncertainty in the measurements,
  including spatial variations in the radiation field and misestimates
  of the continuum.  Variations in the background may contribute to
  scatter in abundance estimates at the $\sim 0.2$ dex level, but do
  not contribute a significant systematic error.  The same is true for
  continuum errors.  If systematic errors are present they would tend
  to enhance, rather than diminish, the evolutionary signal.  The one
  exception to this rule is the proposed ``sawtooth'' shaped
  attenuation of the 3-4 Ryd background from \heii absorption.
  Absorption at the levels indicated in \citet{HM_sawtooth} lead to a
  systematic increase of 0.06 dex in our abundances.  A substantially
  stronger absorption signature could weaken the detected evolution
  signal, but this may also cause conflict with measurements of the
  \siiv/\civ ratio in the limited cases where this could be measured.}
\item{The total carbon contribution to closure density in our fiducial
  model is $\Omega_C\sim 2.7\times 10^{-8}$.  The same quantity
  calculated at $z\sim 2$ is 1.7 times larger, implying that roughly
  half of the heavy elements seen in the \lya forest at $z\sim 2.4$
  were distributed into the IGM in the 1.3 Gyr of the $z\sim 3-4$
  interval.}
\item{The mass flux of carbon into the IGM needed to sustain this
  growth is comparable in magnitude to the carbon yield inferred from
  newly formed stellar populations at these redshifts.  We estimate
  that on global scales (and within very large uncertainties), the
  metal feedback rate from galaxies is around $\sim 30\%$ of the star
  formation rate.  This large rate is consistent with observations of
  star forming galaxies in the local universe.}
\end{enumerate}

The main result of the paper---our detection of a decline in abundance
at higher redshift---shows that at least some of the enrichment of the
IGM is taking place at the time when we observe it.  High redshift
\civ observations show that some metals already existed at $z\sim
5.5$, but a substantial fraction of the heavy elements seen at $z\sim
2.4$ must have been ejected from galaxies we can readily observe.

The transport of metals from galaxies to the IGM is one of the
strongest lines of evidence supporting the notion that feedback is a
crucial element of galaxy formation.  This has long been inferred
indirectly, based on observations and models of galaxy properties in
the local universe.  Our measurements suggest that observations at
high redshift may start to place meaningful constraints on the
feedback process during the epoch when it was actually taking place.

\acknowledgements It is a pleasure to thank the staff of the Magellan
telescopes for their assistance in obtaining the data contained here.
George Becker kindly assisted in reducing much of the data taken with
the MIKE spectrograph.  Francesco Haardt supplied private versions of
the UV background models used, and Claude-Andre Faucher-Giguere kindly
supplied his new models as well and offered several useful
suggestions.  Steve Furlanetto deserves credit for pointing out the
possibility of using \heii methods to study fluctuations in the \civ
background.  Finally, I wish to acknowledge financial support from the
Alfred P. Sloan foundation, and the NSF under grant AST-0908920.  I
also gratefully acknowledge generous lumbar support from the
Adam J. Burgasser Chair in Astrophysics.

\bibliography{z4civ}

\begin{deluxetable}{c c c c c c c c}
\tablewidth{0pc}
\tablecaption{Line Sample \hi and \civ Measurements}

\tablehead{{$z_\mhi$} & {$dz$} & {$b_\mhi$} & {$db$} & {$\nhi$} & {$d\nhi$} & {$N_\mciv$} & {$dN_\mciv$}}
\startdata

\multicolumn{8}{c}{BR0353-3820} \\
\hline
4.474085 & 0.000016 & 23.36 &  0.93 & 15.423 &  0.078 & $\le$1.167450e+12 & \nodata \\
4.460953 & 0.000019 & 20.53 &  2.66 & 15.356 &  0.232 & $\le$2.569830e+12 & \nodata \\
4.454443 & 0.000176 & 32.72 &  4.83 & 15.231 &  0.186 & $\le$4.784500e+12 & \nodata \\
4.449189 & 0.000066 & 42.75 &  1.85 & 14.706 &  0.059 & $\le$1.067320e+12 & \nodata \\
4.440628 & 0.000008 & 25.88 &  0.51 & 15.391 &  0.053 & $\le$2.393390e+12 & \nodata \\
4.436703 & 0.000022 & 20.74 &  1.40 & 14.792 &  0.039 & $\le$1.175590e+12 & \nodata \\
4.428550 & 0.000048 & 25.50 &  6.41 & 14.721 &  0.210 & $\le$1.521730e+12 & \nodata \\
4.427306 & 0.000053 & 23.08 &  4.50 & 14.588 &  0.182 & 3.175900e+12 & 5.549000e+11 \\
 4.387300 & 0.000023 & 27.53 &  0.86 & 15.016 &  0.030 & 1.453000e+12 & 3.534400e+11 \\
 4.365353 & 0.000028 & 26.82 &  2.47 & 14.979 &  0.056 & 4.482700e+12 & 1.404300e+12 \\
 4.363416 & 0.000027 & 30.34 &  2.41 & 15.810 &  0.017 & 6.620800e+12 & 1.705800e+12 \\
 4.362077 & 0.000001 & 41.84 &  2.39 & 15.129 &  0.030 & 6.537800e+12 & 1.692900e+12 \\
 4.358888 & 0.000098 & 49.68 &  2.09 & 16.601 &  0.086 & 1.406300e+13 & 1.466300e+12 \\
 4.357327 & 0.000045 & 37.50 &  4.37 & 15.950 &  0.039 & 1.499900e+13 & 2.223700e+12 \\
 4.355646 & 0.000039 & 42.65 &  1.00 & 15.898 &  0.038 & 9.965700e+12 & 9.962500e+11 \\
 4.347564 & 0.000036 & 22.04 &  2.38 & 15.015 &  0.057 & 9.183500e+12 & 1.373000e+12 \\
 4.349267 & 0.000027 & 39.98 &  4.95 & 16.390 &  0.219 & 9.769600e+12 & 1.266500e+12 \\
 4.345974 & 0.000007 & 25.19 &  0.39 & 14.877 &  0.025 & $\le$2.154060e+12 & \nodata \\
%4.321407 & 0.000127 & 49.20 &  3.67 & 15.572 &  0.083 & $\le$3.810600e+12 & \nodata \\
%4.320024 & 0.000000 & 43.28 &  2.31 & 16.145 &  0.091 & $\le$8.013300e+12 & \nodata \\
4.321407 & 0.000127 & 49.20 &  3.67 & 15.572 &  0.083 & 1.916900e+13 & 1.270200e+12 \\
4.320024 & 0.000000 & 43.28 &  2.31 & 16.145 &  0.091 & 6.301600e+13 & 2.671100e+12 \\
4.317857 & 0.000028 & 23.37 &  3.56 & 14.758 &  0.176 & $\le$1.935430e+12 & \nodata \\
4.312235 & 0.000006 & 24.84 &  1.16 & 14.610 &  0.073 & $\le$1.957310e+12 & \nodata \\
4.283492 & 0.000007 & 22.66 &  0.66 & 15.313 &  0.056 & $\le$9.478000e+11 & \nodata \\
4.279064 & 0.000460 & 29.93 &  4.83 & 14.599 &  1.884 & $\le$7.070800e+11 & \nodata \\
4.275294 & 0.000028 & 29.63 &  1.73 & 15.189 &  0.053 & $\le$9.456400e+11 & \nodata \\
4.270005 & 0.000004 & 26.80 &  0.52 & 15.552 &  0.057 & $\le$1.662370e+12 & \nodata \\
4.266536 & 0.000093 & 20.03 &  7.51 & 14.912 &  0.365 & $\le$1.636960e+12 & \nodata \\
4.265502 & 0.000250 & 17.01 & 12.58 & 14.652 &  0.484 & $\le$2.184220e+12 & \nodata \\
4.262394 & 0.000010 & 25.89 &  1.45 & 14.637 &  0.045 & $\le$5.307900e+12 & \nodata \\
4.251741 & 0.000005 & 18.25 &  0.83 & 14.509 &  0.067 & $\le$1.670080e+12 & \nodata \\
4.248899 & 0.000071 & 38.35 &  2.15 & 15.081 &  0.057 & $\le$1.354510e+12 & \nodata \\
4.246332 & 0.000010 & 25.62 &  0.54 & 14.821 &  0.017 & $\le$2.489940e+12 & \nodata \\
4.215746 & 0.000026 & 37.14 &  1.10 & 15.069 &  0.023 & $\le$1.363900e+12 & \nodata \\
4.214093 & 0.000054 & 35.74 &  2.14 & 14.524 &  0.066 & 3.027200e+12 & 4.501900e+11 \\
 4.211229 & 0.000044 & 43.14 &  3.19 & 15.521 &  0.095 & 2.055800e+13 & 2.462300e+12 \\
 4.207951 & 0.000034 & 38.72 &  2.55 & 15.050 &  0.040 & 1.854900e+12 & 5.446300e+11 \\
 4.203494 & 0.000011 & 20.59 &  0.91 & 14.538 &  0.018 & $\le$2.748370e+12 & \nodata \\
4.195487 & 0.000021 & 38.94 &  3.37 & 16.121 &  0.239 & 1.298400e+13 & 1.459910e+11 \\
 4.187106 & 0.000008 & 60.09 &  1.08 & 15.146 &  0.032 & $\le$2.299030e+12 & \nodata \\
4.181189 & 0.000317 & 43.20 &  9.22 & 14.605 &  0.357 & $\le$2.430070e+12 & \nodata \\
4.179194 & 0.000020 & 31.28 &  1.32 & 15.106 &  0.025 & $\le$3.316110e+12 & \nodata \\
4.159921 & 0.000068 & 48.62 &  6.29 & 14.940 &  0.046 & $\le$6.238000e+11 & \nodata \\
4.157869 & 0.000161 & 57.96 & 19.39 & 14.547 &  0.127 & $\le$7.426000e+11 & \nodata \\
4.156232 & 0.000063 & 22.15 &  5.40 & 15.194 &  0.196 & 2.136000e+12 & 3.455100e+11 \\
 4.150681 & 0.000009 & 24.02 &  0.83 & 15.046 &  0.070 & 6.060600e+11 & 1.882100e+11 \\
 4.128559 & 0.000006 & 25.43 &  0.87 & 15.121 &  0.089 & $\le$9.137800e+11 & \nodata \\
4.084159 & 0.000007 & 28.10 &  1.87 & 15.369 &  0.155 & 6.524400e+12 & 1.244600e+12 \\
 4.065986 & 0.000011 & 20.46 &  0.76 & 14.752 &  0.033 & 2.401500e+12 & 7.137400e+11 \\
 4.061522 & 0.000054 & 20.45 &  2.01 & 14.715 &  0.109 & $\le$1.363140e+12 & \nodata \\
4.060768 & 0.000109 & 29.66 & 13.62 & 14.517 &  0.202 & $\le$7.458400e+11 & \nodata \\
4.058138 & 0.000049 & 41.92 &  6.98 & 14.858 &  0.053 & $\le$2.221440e+12 & \nodata \\
4.056497 & 0.000082 & 35.69 & 11.98 & 14.786 &  0.250 & 2.833900e+12 & 5.575700e+11 \\
 4.048319 & 0.000090 & 45.59 &  9.39 & 16.136 &  0.377 & 1.334600e+13 & 1.830900e+12 \\
 4.039261 & 0.000006 & 29.48 &  0.66 & 14.956 &  0.034 & $\le$3.718200e+12 & \nodata \\
4.024303 & 0.000001 & 20.00 &  1.00 & 14.710 &  0.168 & $\le$1.545500e+12 & \nodata \\
\enddata

\end{deluxetable}

\setcounter{table}{1}
\begin{deluxetable}{c c c c c c c c}
\tablewidth{0pc}
\tablecaption{Line Sample \hi and \civ Measurements ({\em Continued})}

\tablehead{{$z_\mhi$} & {$dz$} & {$b_\mhi$} & {$db$} & {$\nhi$} & {$d\nhi$} & {$N_\mciv$} & {$dN_\mciv$}}
\startdata

\multicolumn{8}{c}{BR0418-5726} \\
\hline
4.353709 & 0.000011 & 29.37 &  0.63 & 15.409 &  0.029 & $\le$4.045600e+12 & \nodata \\
4.352097 & 0.000025 & 17.44 &  1.29 & 14.720 &  0.035 & $\le$3.988130e+12 & \nodata \\
4.343351 & 0.000021 & 15.91 &  1.56 & 15.818 &  0.162 & 4.201200e+12 & 6.709000e+11 \\
 4.326431 & 0.000022 & 42.55 &  1.10 & 14.991 &  0.023 & $\le$3.344200e+12 & \nodata \\
4.320692 & 0.000033 & 27.35 &  1.15 & 14.608 &  0.038 & 1.187000e+12 & 3.475000e+11 \\
 4.296004 & 0.000021 & 23.62 &  1.35 & 16.636 &  0.207 & $\le$2.976100e+12 & \nodata \\
4.294459 & 0.000045 & 28.64 &  2.03 & 15.140 &  0.046 & $\le$2.487100e+12 & \nodata \\
4.279348 & 0.000007 & 13.89 &  1.78 & 14.681 &  0.293 & $\le$1.675300e+12 & \nodata \\
4.274486 & 0.000062 & 28.09 &  1.65 & 15.243 &  0.102 & $\le$1.150750e+12 & \nodata \\
4.271221 & 0.000130 & 46.95 &  5.46 & 14.528 &  0.083 & $\le$2.717920e+12 & \nodata \\
4.259756 & 0.000013 & 19.91 &  0.62 & 14.636 &  0.025 & $\le$1.717482e+12 & \nodata \\
4.249553 & 0.000010 & 28.42 &  1.36 & 14.812 &  0.020 & $\le$4.878660e+12 & \nodata \\
4.246544 & 0.000013 & 39.49 &  1.43 & 15.539 &  0.064 & $\le$1.117540e+12 & \nodata \\
4.236839 & 0.000185 & 27.35 &  6.57 & 16.190 &  0.421 & 1.896900e+13 & 5.357000e+12 \\
 4.238187 & 0.001232 & 58.60 & 130.50 & 15.636 &  1.060 & $\le$3.061680e+13 & \nodata \\
4.239286 & 0.000301 & 29.97 & 23.48 & 15.664 &  0.703 & 2.149500e+13 & 2.059100e+12 \\
 4.221174 & 0.000058 & 43.81 &  5.60 & 15.847 &  0.081 & $\le$1.110400e+14 & \nodata \\
4.219525 & 0.000039 & 15.52 &  8.52 & 17.091 &  2.256 & 4.352500e+13 & 2.246000e+12 \\
 4.218226 & 0.000215 & 45.33 & 16.66 & 15.324 &  0.315 & $\le$2.476850e+12 & \nodata \\
4.215030 & 0.000048 & 42.13 &  3.38 & 14.688 &  0.044 & $\le$2.010750e+12 & \nodata \\
4.209963 & 0.000034 & 29.78 &  1.16 & 15.354 &  0.050 & 1.933300e+12 & 4.152700e+11 \\
 4.204677 & 0.000063 & 21.10 &  3.52 & 15.364 &  0.632 & $\le$7.646100e+12 & \nodata \\
4.199868 & 0.000041 & 24.82 &  3.46 & 15.269 &  0.064 & 1.329100e+13 & 1.268400e+12 \\
 4.198389 & 0.000330 & 46.81 & 20.02 & 15.428 &  0.257 & $\le$3.428510e+12 & \nodata \\
4.196863 & 0.000250 & 50.00 &  7.49 & 15.700 &  0.164 & 2.229900e+12 & 7.301700e+11 \\
 4.176287 & 0.000007 & 21.72 &  1.53 & 14.939 &  0.154 & $\le$2.710600e+12 & \nodata \\
4.174260 & 0.000057 & 36.30 &  2.14 & 14.627 &  0.094 & $\le$1.355270e+12 & \nodata \\
4.159184 & 0.000009 & 26.70 &  1.64 & 15.510 &  0.192 & $\le$2.615960e+12 & \nodata \\
4.139674 & 0.000069 & 38.80 &  1.99 & 15.483 &  0.085 & $\le$5.521110e+12 & \nodata \\
4.130618 & 0.000315 & 19.02 &  5.50 & 14.678 &  0.898 & $\le$1.310500e+12 & \nodata \\
4.124536 & 0.000013 & 34.26 &  1.95 & 15.057 &  0.109 & $\le$3.222700e+12 & \nodata \\
4.120722 & 0.000009 & 47.18 &  1.94 & 14.813 &  0.044 & $\le$3.442600e+12 & \nodata \\
4.108080 & 0.000030 & 59.86 &  2.83 & 15.315 &  0.053 & $\le$3.456100e+12 & \nodata \\
4.094620 & 0.000008 & 24.13 &  0.78 & 15.459 &  0.086 & $\le$2.806240e+12 & \nodata \\
4.084853 & 0.000036 & 29.98 &  2.82 & 14.871 &  0.096 & $\le$6.064000e+11 & \nodata \\
4.081859 & 0.000013 & 24.99 &  1.70 & 15.108 &  0.070 & $\le$3.831100e+12 & \nodata \\
4.079143 & 0.000011 & 34.27 &  2.57 & 15.588 &  0.113 & 2.185100e+13 & 1.434600e+12 \\
 4.065194 & 0.000018 & 30.88 &  1.25 & 15.028 &  0.047 & $\le$1.995340e+12 & \nodata \\
4.063509 & 0.000061 & 28.00 &  2.88 & 14.678 &  0.060 & $\le$2.679990e+12 & \nodata \\
4.056032 & 0.000014 & 33.74 &  1.00 & 15.077 &  0.034 & 3.049600e+12 & 6.418300e+11 \\
 4.054350 & 0.000084 & 27.37 &  3.88 & 14.561 &  0.104 & $\le$2.664230e+12 & \nodata \\
4.053611 & 0.000115 & 16.06 & 11.71 & 14.565 &  0.231 & $\le$3.405100e+12 & \nodata \\
4.052538 & 0.000064 & 25.00 &  1.41 & 16.167 &  0.304 & $\le$1.471000e+12 & \nodata \\
4.019813 & 0.000011 & 27.11 &  1.56 & 14.829 &  0.099 & $\le$2.463100e+12 & \nodata \\
4.017804 & 0.000012 & 35.88 &  3.03 & 15.188 &  0.156 & $\le$3.272670e+12 & \nodata \\
4.013256 & 0.000022 & 40.30 &  4.53 & 15.098 &  0.159 & 9.325000e+12 & 1.006000e+12 \\
 4.008585 & 0.000011 & 17.36 &  2.86 & 14.616 &  0.321 & $\le$1.370500e+12 & \nodata \\
4.006144 & 0.000013 & 27.81 &  2.45 & 14.742 &  0.134 & $\le$5.012040e+12 & \nodata \\
4.003678 & 0.000011 & 30.28 &  2.37 & 15.187 &  0.184 & $\le$3.619320e+12 & \nodata \\
\hline
\enddata
\end{deluxetable}

\setcounter{table}{1}
\begin{deluxetable}{c c c c c c c c}
\tablewidth{0pc}
\tablecaption{Line Sample \hi and \civ Measurements ({\em Continued})}

\tablehead{{$z_\mhi$} & {$dz$} & {$b_\mhi$} & {$db$} & {$\nhi$} & {$d\nhi$} & {$N_\mciv$} & {$dN_\mciv$}}
\startdata

\multicolumn{8}{c}{BR0714-6449} \\
\hline
4.397271 & 0.000008 & 28.22 &  0.71 & 15.140 &  0.054 & $\le$2.065630e+12 & \nodata \\
4.394346 & 0.000026 & 46.01 &  1.19 & 15.321 &  0.026 & 6.065800e+12 & 1.274700e+12 \\
 4.391425 & 0.000034 & 12.99 &  5.94 & 16.467 &  1.750 & $\le$1.431670e+12 & \nodata \\
4.388564 & 0.003111 & 47.10 & 64.55 & 15.540 &  3.027 & 1.953600e+13 & 1.720900e+12 \\
 4.383436 & 0.000032 & 45.08 &  2.72 & 15.412 &  0.068 & $\le$1.610860e+12 & \nodata \\
4.381303 & 0.000027 & 30.27 &  1.45 & 15.343 &  0.068 & $\le$1.665610e+12 & \nodata \\
4.376600 & 0.000032 & 48.01 &  1.89 & 14.808 &  0.028 & $\le$2.499250e+12 & \nodata \\
4.371056 & 0.000055 & 25.82 &  2.18 & 15.598 &  0.196 & $\le$1.634500e+12 & \nodata \\
4.363155 & 0.000024 & 48.49 &  1.19 & 15.818 &  0.052 & $\le$5.346400e+12 & \nodata \\
4.359026 & 0.000014 & 38.34 &  1.97 & 14.632 &  0.067 & $\le$3.695800e+12 & \nodata \\
4.345805 & 0.000011 & 28.01 &  1.79 & 14.602 &  0.098 & 3.804600e+12 & 1.078000e+12 \\
 4.342708 & 0.000038 & 29.64 &  3.74 & 14.516 &  0.101 & $\le$2.709610e+12 & \nodata \\
4.339696 & 0.000008 & 16.83 &  0.44 & 14.933 &  0.057 & $\le$2.503270e+12 & \nodata \\
4.333522 & 0.000012 & 12.46 &  2.88 & 15.394 &  0.873 & $\le$3.287530e+12 & \nodata \\
4.330081 & 0.000011 & 32.24 &  0.89 & 15.838 &  0.083 & $\le$2.529160e+12 & \nodata \\
4.326107 & 0.000015 & 30.21 &  1.60 & 15.421 &  0.087 & $\le$2.999980e+12 & \nodata \\
4.323170 & 0.000023 & 32.99 &  1.69 & 15.390 &  0.080 & $\le$3.884960e+12 & \nodata \\
4.318676 & 0.000013 & 21.08 &  0.75 & 14.726 &  0.035 & $\le$7.277800e+12 & \nodata \\
4.290688 & 0.000008 & 18.97 &  1.76 & 15.522 &  0.344 & $\le$3.648910e+12 & \nodata \\
4.272618 & 0.000009 & 19.92 &  1.06 & 15.098 &  0.133 & 2.932500e+12 & 7.360300e+11 \\
 4.269980 & 0.000040 & 21.24 &  1.47 & 14.926 &  0.100 & $\le$8.070100e+11 & \nodata \\
4.261784 & 0.000452 & 38.76 & 24.13 & 14.603 &  0.358 & $\le$4.188790e+12 & \nodata \\
4.243142 & 0.000009 & 18.81 &  1.23 & 15.648 &  0.190 & $\le$2.221580e+12 & \nodata \\
4.233165 & 0.000021 & 20.75 &  1.08 & 15.199 &  0.091 & $\le$3.497500e+12 & \nodata \\
4.213589 & 0.000048 & 32.81 &  3.06 & 15.227 &  0.049 & $\le$2.230290e+12 & \nodata \\
4.200533 & 0.000030 & 44.15 &  2.22 & 14.712 &  0.032 & $\le$2.774050e+12 & \nodata \\
4.197665 & 0.000312 & 42.29 &  6.58 & 16.316 &  0.700 & 1.792100e+14 & 2.178100e+13 \\
 4.187770 & 0.000034 & 26.30 &  1.45 & 14.791 &  0.047 & $\le$2.836570e+12 & \nodata \\
4.185000 & 0.000156 & 70.00 & 13.32 & 15.023 &  0.073 & $\le$9.562500e+12 & \nodata \\
4.183234 & 0.000033 & 33.25 &  2.95 & 15.488 &  0.065 & 7.748100e+12 & 8.997800e+11 \\
 4.181524 & 0.000024 & 23.86 &  2.08 & 15.039 &  0.112 & $\le$1.065910e+12 & \nodata \\
4.169388 & 0.000035 & 41.54 &  1.49 & 15.055 &  0.034 & $\le$2.024560e+12 & \nodata \\
4.159509 & 0.000039 & 22.68 &  4.96 & 16.186 &  1.037 & $\le$5.225800e+12 & \nodata \\
4.155559 & 0.000783 & 25.00 & 23.55 & 15.071 &  1.210 & $\le$2.646090e+12 & \nodata \\
4.150633 & 0.000462 & 31.29 & 18.51 & 15.541 &  1.000 & 5.041800e+12 & 1.463100e+12 \\
 4.149312 & 0.000054 & 20.76 &  3.25 & 15.040 &  0.125 & $\le$1.921510e+12 & \nodata \\
4.143332 & 0.000050 & 48.19 &  3.01 & 15.036 &  0.038 & 2.864000e+12 & 6.603200e+11 \\
 4.139893 & 0.000042 & 70.00 &  5.18 & 16.185 &  0.172 & $\le$1.117390e+12 & \nodata \\
4.128364 & 0.000017 & 30.32 &  2.31 & 15.949 &  0.236 & $\le$1.328410e+12 & \nodata \\
4.122314 & 0.000019 & 18.16 &  3.56 & 16.404 &  0.953 & $\le$2.115940e+12 & \nodata \\

\enddata
\end{deluxetable}

\end{document}